\documentstyle[12pt,aaspp4]{article}
\makeatletter

\@ifundefined{chapter}{\def\thebibliography#1{\section*{References\@mkboth
  {REFERENCES}{REFERENCES}}\list
  {\relax}{\setlength{\labelsep}{0em}
        \setlength{\itemindent}{-\bibhang}
        \setlength{\itemsep}{0pt}
        \setlength{\parsep}{0pt}
        \setlength{\leftmargin}{\bibhang}}
    \def\newblock{\hskip .11em plus .33em minus .07em}
    \sloppy\clubpenalty4000\widowpenalty4000
    \sfcode`\.=1000\relax}}%
{\def\thebibliography#1{\chapter*{Bibliography\@mkboth
  {BIBLIOGRAPHY}{BIBLIOGRAPHY}}\list
  {\relax}{\setlength{\labelsep}{0em}
        \setlength{\itemindent}{-\bibhang}
        \setlength{\itemsep}{0pt}
        \setlength{\parsep}{0pt}
        \setlength{\leftmargin}{\bibhang}}
    \def\newblock{\hskip .11em plus .33em minus .07em}
    \sloppy\clubpenalty4000\widowpenalty4000
    \sfcode`\.=1000\relax}}

\newlength{\bibhang}
\let\@internalcite\cite
\def\cite{\let\@citeleft(\let\@citeright)%
    \@ifstar{\citeyear}{\citefull}}
\def\acite{\let\@citeleft\relax\let\@citeright\relax%
    \@ifstar{\citeyear}{\acitefull}}
\def\citenp{\let\@citeleft\relax\let\@citeright\relax
    \@ifstar{\citeyear}{\citefull}}
\def\citefull{\def\astroncite##1##2{##1~##2}\@internalcite}
\def\citeyear{\def\astroncite##1##2{##2}\@internalcite}
\def\acitefull{\def\astroncite##1##2{##1~(##2)}\@internalcite}

\def\@citex[#1]#2{\if@filesw\immediate\write\@auxout{\string\citation{#2}}\fi
  \def\@citea{}\@cite{\@for\@citeb:=#2\do
    {\@citea\def\@citea{; }\@ifundefined
       {b@\@citeb}{{\bf ?}\@warning
       {Citation `\@citeb' on page \thepage \space undefined}}%
{\csname b@\@citeb\endcsname}}}{#1}}

\def\@cite#1#2{\@citeleft#1\if@tempswa , #2\fi\@citeright}
\def\@biblabel#1{}

\makeatother

\newcommand{\PSbox}[3]{\mbox{\rule{0in}{#3}\includegraphics{#1}\hspace{#2}}}
\newcommand{\FigNum}[1]{\unitlength 1pt \begin{picture}(55,10)(-400,35) 
                        \put(0,0){Figure #1}
                        \end{picture}}
 
\newcommand{\bms}{\mbox{b$\mu$s}} 
 
\newcommand{\persec}{\mbox{$\second^{-1}$}}
\newcommand{\percm}{\mbox{$\cm^{-2}$}}
\newcommand{\ppm}{\mbox{$\pm$}}
\newcommand{\cgsflux}{\erg\percm\persec}

\newcommand\approxlt{\mbox{$^{<}\hspace{-0.24cm}_{\sim}$}}

\newcommand{\ee}[1]{\mbox{$10^{#1}$}}
\newcommand{\tee}[1]{\mbox{$\times 10^{#1}$}}

\def\chisqr{\mbox{$\chi^2$}}
\def\chisqrnu{\mbox{$\chi^2_\nu$}}
\def\x1608{{4U~1608$-$522}}

\def\cenx4{{Cen~X$-$4}}

\def\saxj1808{{SAX J1808.4$-$3658}}

\newcommand{\cm}{\mbox{$\rm\,cm$}}

\newcommand{\second}{\mbox{$\rm\,s$}}

\newcommand{\erg}{\mbox{$\rm\,erg$}}

\newcommand{\rhessi}{{\em RHESSI\/}}

\newcommand{\grb}{{GRB~021206\/}}
\newcommand{\degree}{\mbox{$\, {\rm  deg}$}}
\def\itheta{{\mbox{$D(\theta)$}}}
\def\intheta{{\mbox{$D_{\rm null}(\theta)$}}}

\begin{document}

\title{Re-Analysis of Polarization in the $\gamma$-ray flux of \grb}

\author{Robert E. Rutledge and Derek B. Fox\altaffilmark{1}}
\altaffiltext{1}{Division of Physics, Mathematics and Astronomy; California Institute of
Technology, MS 130-33, Pasadena, CA 91125; rutledge@tapir.caltech.edu,
  derekfox@astro.caltech.edu}

\begin{abstract}
A previous analysis of the {\em Reuven Ramaty} High Energy Solar
Spectroscopic Imager (\rhessi) observation of \grb\ found that the
gamma-ray flux was 80\ppm20\% polarized.  
We re-examine this data and find no signal that
can be interpreted as due to polarization.  First, we find that the
number of scattering events suitable for measuring polarization --
having been scattered from one detector to another, with a count
produced in both -- is considerably lower than estimated by CB03, by a
factor of 10 ($830\pm 150$, vs.\ $9840\pm 96$).  The signal-to-noise
of the data-set is thus too low to produce a detection, even from a
100\% polarized source.  Nonetheless, we develop a
polarization-detection analysis limited in sensitivity only by Poisson
noise, which does not require a space-craft mass model to detect
polarization, as in CB03.  We find no signal which might be
interpreted as due to polarization of \grb.  Separately, we reproduce
the CB03 signal and show that it is not due to polarization.  Rather,
the CB03 signal is consistent with the previously-neglected systematic
uncertainty in the ``null lightcurve'' used for detection.  Due to the
low signal-to-noise ratio of the \rhessi\ data, our Poisson
noise-limited analysis results in an upper limit consistent with
100\%-polarization of the gamma-ray flux from \grb.  Thus, no
observational constraint on the polarization of \grb\ can be derived
from these data.

\end{abstract}

\section{Introduction}

Gamma-ray bursts (GRBs) are associated with the deaths of massive
stars, as seen first in their association with blue galaxies
\cite{bloom99,bloom02b}; the appearance of supernovae explosion (SNe) -like
lightcurves at late times \cite{bloom02}; and the recent convincing
observation of a SNe spectrum associated with a GRB afterglow
\cite{hjorth03,stanek03}.  However, what differentiates GRBs from the 10$^4$
more frequently observed core-collapse, or type-II,  SNe remains uncertain. 

Recently, it was reported \cite[CB03 hereafter]{coburn03} that
80\ppm20\% polarization in the prompt, or burst, gamma-rays of
GRB021206 was detected using the Spectroscopic Imager \cite{spex} on
board the \rhessi\ satellite \cite{rhessi}.  Although not designed
for gamma-ray polarization measurements of GRBs, the multiple
detectors of \rhessi\ were used to search for simultaneous events in
two detectors due to Compton scattering, which then give a preferred
scattering direction projected on the sky.  An excess of scattered
events in a particular direction on the sky was interpreted as due to
angle-dependent scattering associated with intrinsic polarization of
the gamma-ray photons.

Such a high polarization fraction in the gamma-rays may require a
large-scale persistent magnetic field -- of a strength observed thus
far only from neutron stars, in particular the magnetars which are
believed to account for \approxlt 10\% of the observed NS population.
If this property of GRBs is confirmed in future observations, then
polarization would strongly constrain GRB production models, and may
provide important input physics to understanding the generalized SNe
phenomena.  The reported detection has led to the wide discussion of
mechanisms for producing the high polarization, which would constrain
the emission and progenitor models
\cite{lyutikov03,granot03,eichler03,nakar03,lazzati03,matsumiya03},
and has prompted discussion of detector development to better observe
gamma-ray polarization \cite{bloser03}.

The analysis of CB03 used a mass-model for the \rhessi\ satellite to
calculate the satellite response to an unpolarized beam, in order to
calculate the unpolarized lightcurve.  It is first needed to calculate
the "null" lightcurve (that is, the lightcurve of an unpolarized GRB)
to demonstrate detection.  It is also needed to correct the detected
signal, and dominates the error bar in the measurement of 80\ppm20\%
polarization.

We describe the observation in \S\ref{sec:obs}, and give an overview
of the analysis by CB03 in \S\ref{sec:cb03}.  To produce the
polarization measurement, it is necessary that photons scattered
between two detectors be observed in the data, and in
\S\ref{sec:double} we estimate the number of observed scattered
double-count events, finding it to be a factor of 10 lower than the
previous estimate.  In \S\ref{sec:singlecounts}, we examine the
distribution of single counts and double counts as a function of the
instantaneous position of the detectors where the counts are detected,
and find that, while both single- and double- counts exhibit strong
dependencies on this position, their ratio does not, suggesting the
absence of a signal due to polarization.  We nonetheless describe in
\S\ref{sec:anal} a polarization detection analysis suitable for
\rhessi\ data which is limited in its sensitivity by Poisson
statistics (that is, not dependent upon the space-craft mass model)
and apply it to the observation of \grb, deriving a correction which
accounts for the angle-dependent Klein-Nishina cross section in
\S\ref{sec:kn}.  The results of the polarization analysis are
discussed in \S\ref{sec:results}, finding no evidence for a signal
which could be due to polarized gamma-ray photons from \grb.  In
\S\ref{sec:bad}, we reproduce the modulation reported by CB03, finding
that the reported signal is not observed using the new method, which
does not rely upon the Monte Carlo radiative transfer through the
\rhessi\ mass model.  In \S\ref{sec:con} we summarize our results,
compare them with those of CB03 and conclude.

\subsection{Observation of \grb\  with \rhessi}
\label{sec:obs}

The observation of \grb\ is described by CB03.  We recount pertinent
information.  The \rhessi\ satellite is a gamma-ray imager primarily
for solar observations using a rotation modulated collimator.  It is
in a low-Earth orbit, with its imaging (z) axis staring within a few
arc-minutes of the solar center.  The entire satellite rotates
approximately around this $z$ axis clockwise from north, with a period
of $\sim$4 seconds.

The \rhessi\ spectrometer detectors are an array of 9 separate,
effectively identical detectors, each a cylinder roughly 7.1\, cm
diameter.  Each detector is electronically separated into two segments
(``front'' and ``rear'') which have slightly different photon energy
responses. A diagram of the detectors and their orientation within the
spacecraft detector plane is given in \cite{mcconnell02}.  We give a
similar diagram in Fig.~\ref{fig:spex}, and the mass-model coordinate
centers of the detectors in Table~\ref{tab:spex}.

The approximate start time of GRB021206 measured with \rhessi\ was
2002-Dec-06 22:49:16 UT.  The best gamma-ray localization of GRB021206
was found \cite{hurley03} to be an error ellipse centered at
R.A.=240.195 deg (16h00m46.8s), dec.=$-$9.710 deg ($-$09d42m36s)(J2000),
with a major axis of 20.4\arcmin and a minor axis of 0.53\arcmin,
position angle of -18\degree\  (relative to the positive direction of
declination).  We adopt the position of a related radio transient
\cite{frail03}.  

At the time of the transient, the solar location was about
R.A.=16h53m20.15s, dec.=-22d32m47.7s. The GRB was localized at an
angle $\theta_{\rm GRB}=45$\degree\ clockwise from N of the sun, and
offset from the direction of solar center by 17.98\degree.  Counts
detected with the \rhessi\ spectrometer are timestamped with
resolution of $2^{-20}$\, s, or 1 binary micro-sec (\bms); the
detector segment; and photon energy.

\subsection{Approach of CB03}
\label{sec:cb03}
We refer the reader to CB03 for details of the analysis used to
detect the polarization signal.  We describe the pertinent approach
here. 

A double-count scattering event takes place when a gamma-ray photon
makes a single scatter off of detector $i$, producing a count in that
detector, then leaving that detector to be partially or fully absorbed
in a second detector $j$, producing a simultaneous count in detector
$j$.  A double-count coincidence  event, which appears identical to a
double-count scattering event, takes place when two unrelated photons
produce simultaneous counts in two different detectors.  A
single-count event takes place when a single photon is fully or
partially absorbed in a detector $i$, but is not followed by a
detection in another detector within some period of time $\Delta T$ as
measured by the on-board clock.

When a double-count event is registered, the position angle on the sky
($\theta$) of the line joining the centers of the detector pair is
determined.  From this, a lightcurve of double count events as a function
of $\theta$ was produced.

CB03 produced a null double-count event lightcurve to be subtracted
from the observed lightcurve by Monte Carlo simulation (S. Boggs,
W. Coburn, priv. comm.).  They simulated $\sim$18,000,000 photons
propagating through the spacecraft from the direction of the GRB, to
produce a library of double-event scatters between detectors.

Following this subtraction, a sinusoidal modulation in the difference
lightcurve was found, which was significantly different from a
constant value.  Based on this sinusoidal residual, CB03 concluded
that the incoming gamma-ray beam was polarized.  Using the scattering
fractions found with the GEANT mass model, the magnitude of the
modulation was corrected to find the intrinsic polarization magnitude
of 80\ppm20\%.  The uncertainty is dominated by uncertainty in the
GEANT mass model.

\section{Proportion of Scattering Events and Backgrounds}
\label{sec:double} 
\newcommand{\bus}{\bms}

To estimate the sensitivity of the \rhessi\ spectrometer to gamma-ray
polarization, it is critical to establish the presence and estimate
the number of two-detector scattering events in the data, and to
minimize, and also estimate, the level of contaminating two-detector
backgrounds.

It is possible to analyze the \rhessi\ data for evidence of a
polarization signature without investigating these effects.  In most
but not all cases (see \S6), double-count event backgrounds will
dilute the angular modulation signal but will not generate an
angularly-modulated signal of their own, and thus will not mimic the
effects of genuine gamma-ray polarization.  However, even when
modulation can be detected in this fashion, estimating the number of
two-detector scatters and associated backgrounds still allows a
valuable consistency check of the analysis to be made.  Moreover,
converting a detection of modulation to an estimate of the
polarization signal strength requires an accurate estimate of the
signal and background.

CB03 present estimates of the number of two-detector scatter events
and backgrounds in their ``Methods'' section.  They find $9840\pm 96$
two-detector scatters, and estimate a background of $4488\pm 72$
two-detector coincidences and $588\pm 24$ ``background scatter''
events, for a total of 14916 double-count events detected.  The
meaning of the ``background scatter'' category is not defined but
would seem to refer to scattering events attributable to background
sources other than \grb\ itself.  They also state that the number of
two-detector scatter events is ``roughly 10\% of the total
0.15--2.0~MeV light-curve events,'' and that this proportion is in
agreement with the results from their Monte Carlo simulations; the
nature and degree of this agreement is not quantified.

In this section we make a model-independent investigation of the
fraction of two-detector scatter events in the \rhessi\ data for
\grb.  In particular, we determine the number of two-detector events
which are due to events of each of the following types: (a) scattered
events, which produce two counts in two different detectors from a
single scattered photon; (b) the coincident arrival of two unrelated
photons in two different detectors; (c) scattered events, which
produce two events in the same detector; and (d) events which take
place in different detectors due to a detector effect.  For the
polarization measurement, events of type (a) permit measurement of
polarization; type (b) is the irreducible background; and types (c)
and (d) are background which, in principle, can be identified by their
non-Poissonian nature and removed from the data.


\subsection{Relative Timing Accuracy of \rhessi\  Detectors}

As a first step, we examine the timing properties of \rhessi\
spectrometer data to establish an appropriate definition for selection
of ``simultaneous'' events.  We find that it is necessary to accept
events within a time window of 5\,\bus\ to avoid discarding
simultaneous events.

Fig.~\ref{fig:diffdt}a is a histogram of wait times between timestamps
of consecutive counts ($\Delta T$); for a pure Poisson process -- not
the present case, because of the strong variations in the GRB mean
count rate -- this histogram will exhibit the classical exponential
form.  We observe a clear non-Poissonian excess near $\Delta T\approx
2\,\bms$, that is, at $\Delta T$ greater than zero. Since the
light-travel time for 2\,\bms\ is 570\,m, far larger than the
separations between detectors, this cannot be a physical effect,
but rather must be due to behavior of the detector electronics, with
simultaneous events not receiving identical timestamps.

In Fig.~\ref{fig:diffdt}b we have removed the 6732 event pairs which
take place either simultaneously or consecutively (with no intervening
counts) in the same detector; the front and rear segments of a
detector are considered to be the same detector for these purposes.
Note that double-count, same-detector events are not useful for
polarization measurement.  While these events are a significant
fraction of those with $\Delta T\leq5$, we still observe a
non-Poissonian excess of counts at $\Delta T$=1--4\,\bms; in addition,
we can see non-Poissonian excesses at $\Delta T$=8, 12, 16, 20, 24, 28
and 32\, \bms\footnote{The excesses are even more apparent if one
looks only at events detected in the rear segments.  The origin of
these excess counts at 4\,\bms\ is related to a detector deadtime
effect (D. Smith, priv. comm.)}.  These excesses indicate that the
spacecraft timestamps are subject to systematic uncertainties at the
$\Delta T \sim 4\,\bms$ level, and possibly even at the 32\,\bms\
level.  Thus, we treat event pairs with
$\Delta T \leq 5\,\bms$ as simultaneous in our subsequent analyses.

As a consequence of this definition, if we observe more than two
consecutive counts within 5\,\bms\ of each other, their
simultaneity  makes it impossible to extract a unique scattering
event from the group.  We therefore exclude all events within such
multi-event groupings from our analysis.


\subsection{Double-Count Events Due to Coincidence}
\label{sec:coincidence}
To evaluate the degree to which the counts statistics are Poissonian,
we bin the data into a lightcurve with time resolution $\Delta
T$=5\,\bus.
 
To account for the highly variable nature of the GRB light curve, we
estimate the countrate in $\delta t = 0.005$\,s intervals, treating
the count rate over each interval $i$ as constant for purposes of
estimating the Poissonian mean rate $\mu_i$.  For the entire selected
portion of the GRB light curve, we find the number $N_{\rm j, obs}$ of
5\,\bms\ bins containing $j$ counts to be $N_{\rm 0, obs.}$=969787,
$N_{\rm 1, obs.}$=72510, $N_{\rm 2, obs.}$=5666, and $N_{\rm >2,
obs.}$=481 bins, respectively.

Estimating the Poisson mean rate $\mu_i=N_{{\rm 1, obs}, i}/N_{{\rm 0,
obs}, i}$ in each interval from the number of single-event and
zero-event time bins, we calculate the number of 5~\bus\ time bins
within that interval that will contain 2 ($N_2$) or $>$2
($N_{>2}$) events, respectively, due to Poisson fluctuations:
\begin{eqnarray}
    N_2 &=& \frac{\delta t}{\Delta T} \sum_i \, \frac{\mu_i^2}{2} e^{-\mu_i}\\
    N_{>2} &=& \frac{\delta t}{\Delta T} \sum_i 1 - (1 + \mu_i +
               \frac{\mu_i^2}{2}) e^{-\mu_i}
\end{eqnarray}
Summing over all $\delta t$ intervals, we find $N_2=3760\pm 30$\,bins.
The remaining excess of $(5666-3760) = 1906\pm 81$ double-event bins
are not due to coincidence (event type b), and must be due to
non-Poisson processes (event types a, c, or d).

A disproportionate fraction of these arise from double-count events
within a single detector (event type c).  Of the $N_{\rm 2, obs} =
5666$ events, 1313 are due to simultaneous events in the same
detector, compared to the $1/9\times 5666\approx 630$ events
expected from coincidence.

As a result, when we require the two events in each double-event bin
to rise from different detectors, we find an observed number of
$N_{\rm 2, obs, dd} = 4983\pm 83$ double-event, different-detector
bins compared to an estimated number of $N_{\rm 2, dd} = 8/9\times N_2
= 3342\pm 27$ due to Poisson coincidence alone.  This implies that
$1641\pm 87$ double-event, different-detector bins may be due to
scattering or unidentified detector effects (event types a or d).


\subsection{Double- and Multi-Count Events Due to an Unknown Process}

While we find the number of bins with $>2$ events due to coincidence
is $N_{>2}$=159\ppm 2 bins, we observe $N_{\rm >2, obs.}$=481~bins,
for an excess of 322\ppm22 bins, which must be due to non-Poisson
processes (event types a, c or d).

If these non-Poisson events were due exclusively to scattering
processes, then the number of double-count events will be a fraction
$f$ of single events $N_{\rm 2, scattering}=f\, N_1$, the number of
triple scattering events $N_{\rm 3, scattering}=f^2\, N_1$, and number
of $j$- scattering events $N_{\rm j, scattering}=f^{j-1}\, N_1$.
Thus, the ratio $r=N_{\rm 2, scattering}/N_{\rm >2, scattering} =
1./(\sum_{i=1} {f^i})$.  Taking $f=N_{\rm 2, obs.}/N_{1, obs.} =
1906/85387$=(2.2\ppm0.1)\tee{-2}, we find $r=44\pm2$, while we observe
$r_{\rm obs.}=N_{\rm 2, obs.}/N_{\rm >2, obs.}=1906/322=5.9\ppm0.5$
(here, we have not corrected for double-count events due to scatters
within a detector)\footnote{The cross section for Compton scattering
increases by a factor of $\sim$2 between 1\, MeV and 150\, keV.  In a
worse-case, the $N_{\rm 3, scattering}=(2f)fN_1$ and $N_{\rm N>2,
scattering}=(2f)^{N-1}fN_1$, and $f=22.7$ -- still discrepant with the
observations.}.

Thus, there are a factor 7.8\ppm0.8 more $N_{\rm >2, obs.}$ events
relative to the number of $N_2$ events than can be explained by
scattering events.  We compared the distribution of detectors and
photon energy for counts included among the observed $N_{>2}$ events,
and found they do not differ significantly from the same distributions
of all the GRB data.  Specifically, the multi-count events typically
occur in more than one detector.

We performed an identical analysis on data from during a background
period (43180 counts in 24\, s), and found non-Poissonian $N_{>2}$
bins present in those data as well. The background period should have
produced $N_{\rm >2, bg}=0.2$ bins (on average), but $N_{\rm 2, bg,
obs.}$=1013.  Also during this period, $N_{\rm 2, bg}=105\pm1$, while
$N_{\rm 2, bg, obs.}$=3827 bins were observed.  Therefore, these
non-Poissonian excess events are not associated with the GRB.  

We attribute this non-Poissonian excess to an unknown process --
perhaps particle background, or noise in the detector electronics or
instrumentation -- which we are unable to precisely model.
Nonetheless, based on the background period, the non-Poisson process
produces a ratio of counts $r_{\rm obs.,\, bg}= (3827-105)/(1013-0.2)
= 3.7\pm0.1$, significantly below the value of 5.9\ppm0.5 observed
during the burst.

It seems reasonable that the same unknown process can also produce
double-count events, contributing to the double-count event
background; however, the process is clearly non-Poissonian, and we are
unable to produce a demonstrably reliable estimate of the number of
such events\footnote{If, however, we assume that $r_{\rm
obs.}$ is constant between the background and the GRB observation,
then the number of double-count events during the GRB which could be
due to type a events is 1906\ppm81 - 3.7\ppm0.1$\times$ (322\ppm22) =
715\ppm120 counts.  }.  Nonetheless, we note the effect and its
uncertain contribution to the irreducible background as an event type
d.
\label{sec:percent}

\subsection{Summary of Corrections for Double-Count Event Rate}

From analysis in this section, we conclude: 

\begin{enumerate}
\item counts with time-tags  different by $\leq$ 5 \bms\ should be
  regarded as simultaneous; 
\item double-count events within a single detector contribute
significantly, adding to the background of double-count events, but
are easily identified in the data stream, and should  be
replaced with a single count at the earlier time-stamp; 
\item An unknown  effect produces multi-count
  groups ($> 2$) of simultaneous counts in different detectors
  in excess of the number expected from coincidence,  and
  which cannot be attributed to a scattering process, both during the
  GRB and during an earlier background period.  These groups can be
  identified in the data-stream as occurring in a period $\leq$5\bms;
  since they cannot be used for measuring polarization -- as which two
  of the $>$ 2 events were the sequentially first cannot be
  determined -- these events should be removed prior to analysis; 
\item The same non-Poissonian, non-scattering effect likely also
  produces double-count events; we can estimate the number of
  double-count events based on the ratio $r_{\rm obs.}=(N_{\rm 2,
  obs.}-N_2)/(N_{\rm >2, obs.} - N_{>2})$
  (correcting each value for the coincidence rate) observed during the
  background observation.  However since we cannot model the
  non-Poissonian process which is responsible for this ratio, the
  ratio has an unquantifiable systematic  uncertainty. 
\end{enumerate}

Based on this analysis, we sequentially apply the following selections
to the GRB data to obtain double-count events using un-binned data:

\begin{itemize}
\item Replace all double-count events in the same detector with
  single-counts, at the earlier time stamp (83300 counts remaining
  afterwards).  This removes ``echo'' counts in the detector
  \cite{smithself02}. 
\item Remove  groups of $>$2 events which take place in $\leq$5\, \bms,
($N_{\rm >2, obs.}=$719 such multi-count events, which is in excess of the $N_{>2}$=474 such events expected from Poisson statistics, leaving  81034 counts).
\end{itemize}

Following these selections, we have $N_{\rm 2, obs.}$=8230
double-count events remaining. This is fewer than the $N_{\rm 2,
obs.}=14916$ found by CB03.  We list the contributions of ``signal''
and the irreducible background to these totals in Table~\ref{tab:bg}.

To estimate the contribution to the irreducible background caused by
coincidence of two unrelated photons in two different detectors, we
use a similar approach as in \S~\ref{sec:coincidence}.  We use an
integration time of $\delta t=$0.005 sec, and find the number of
single counts which have another which follows within $\Delta
T=5$\bms.  To do so, we calculate $N_n$, the number of unbinned events
which are a collection of $n$ counts within a period $\Delta T$: 

$$
N_n = \sum_i \frac{\mu_i^{n-1}}{(n-1)!} \exp(-\mu_i) (N_i -(n-1))
$$

\noindent where $N_i$ is the number of detected counts in bin $i$, and
$\mu_i=N_i \Delta T/\delta t$.  Using this we find $N_2$=6640\ppm80,
after correcting by a factor 8/9 to account for the requirement that
the counts be in two different detectors.  This is greater than that
found by CB03 ($N_{2}$=4488\ppm72). If we were to assume a constant
countrate throughout the burst we obtain $N_{2}=5380$ -- lower than
what we find for a variable intensity burst, but above that of CB03. 

To estimate the contribution to the irreducible background due to the
``unknown'' process, we performed the same analysis as in
\S\ref{sec:percent}, but to the unbinned data, obtaining $r_{\rm
obs.}$=3.1\ppm0.1 from the background period, giving us $N_{\rm 2,
bkg}=r_{\rm obs.}\times(N_{>2, obs.}-N_{>2})=
3.1\pm0.1(719-474)=760\ppm110$ double-count events due to the unknown
process (this value has an unquantifiable systematic uncertainty).

The result of this is that we find significantly fewer non-coincidence
double-count events than found by CB03 (830\ppm150 double-count
events, vs. 9840\ppm96 by CB03).

We use these single and double-count event lists for analyses in
\S\ref{sec:simple}, \S\ref{sec:singlecounts} \& \S\ref{sec:anal}.
\label{sec:dcounts}

\subsection{A Simpler Method of Determining the
  Double-Count Event Rate due to Scattering Between Detectors}
\label{sec:simple}

There is a method to estimate the fraction of double-count events
which are due to scattering between detectors considerably more
straightforward than attempting to identify all the backgrounds, as we do
above. 

First, we note that for each  detector the total solid angle
subtended by secondary detectors varies (see Fig.~\ref{fig:spex}).  In
particular, the relative solid angle seen by the inner detectors
(numbers 1, 2, and 7) is significantly larger than that seen by the
outer detectors (numbers 3--6, 8, and 9), and if scattering events
represent a significant fraction of the total then this should be
reflected in the number of double-count events involving each
detector.

For the analysis in this sub-section only, we exclude detector 2, due
to its different electrical set-up.  We calculate, for each detector,
the total physical solid angle ($\Omega_i$) subtended by the other
seven detectors, except where the lines of sight to the other
detectors are blocked by  detector 2.  We also examined
the total number of single-counts in each detector $I_i$ observed
during the GRB -- if all detectors were equally sensitive (including
shadowing by the spacecraft), they should have equal values of $I_i$,
but they do not, indicating different sensitivities during the GRB.
We use $I_i$ as a measure of the detector responses (see
Table~\ref{tab:spex} for $\Omega_i$ and $I_i$; note that we use the
counts selected as described in \S\ref{sec:dcounts}).

We then determine $N_{\rm 2, obs}(i)$ -- the number of double-count events in
which one of the counts appears in detector $i$. We model this
observed event rate with the function: 

\begin{equation}
N_2(i) = N_{\rm 2, tot} \left(\frac{I_i\Omega_i}{\sum_j I_j\Omega_j} f + 
\frac{I_i}{\sum_j I_j}(1-f)\right)
\end{equation}

\noindent where $N_{\rm 2, tot}$ is held fixed at the observed number of
counts for the double-events in detectors 1 and 3-9 (10906).  Here,
the coefficient on $f$ is due to the scattered events involving two
detectors.  The coefficient on $1-f$ is the random background.  The
value $f$ can be measured because the values of $\Omega_i$ are
different for different detectors.  Our analysis ignores two effects:
the angular dependence of the Klein-Nishina cross section for
scattering, and the absorption/scattering effects of passive material
between detector units, which will decrease the effective $\Omega_i$
by varying amounts, depending on the distance between detectors. 

Using a \chisqr\ minimization fitting technique, we obtain an
acceptable fit to the data (\chisqr=9.28 for 7 degrees of freedom --
dof), for $f=0.11$\ppm0.03.  The best fit and residuals, including the
proportion of double events associated with scattering and
coincidence, respectively, for each detector, are shown in
Figure~\ref{fig:derekfit}.  Therefore, we find 11\ppm3\% of the
double-count events are due to scattering between detectors.

Comparing this with the value we found by attempting to eliminate all
the reducible background sources we find (11\ppm3\%$\times$8230\ppm90
double-count events)=910\ppm250 double-count events due to scattering
between detectors (where here use double-count events in all
detectors, including detector 2).  This is in agreement with the value
of 830\ppm130 double-count events we found from  the detailed analysis of
the different background contributions.

\section{Distribution of Observed Counts with Detector Angle}
\label{sec:singlecounts}

Fig.~\ref{fig:ddsolid}a presents the number of single counts $S$ as a
function of detector angle $\theta$, where for each event $\theta$
indicates the position of the detector at the time of the event,
relative to the center of the detector array in a non-rotating
celestial frame ($\theta=0$ is celestial north). 

We see that there is a clear bias for single events to occur towards
one side of the spacecraft.  This could be due to a number of effects.
For example, different detector sensitivities produce uneven exposure
as a function of time over the range of $\theta$ which,  combined
with the GRB intensity variability (Fig.~\ref{fig:lc}),  can produce
such a bias.  Also, Earth-scattered gamma-rays -- which may constitute
a significant fraction of the total events -- will preferentially
illuminate the side of the spacecraft towards the Earth (see
\citenp{mcconnell02} for discussion).

The increased rate of single events towards one side of the spacecraft
will produce an increased rate of double events on the same side
through the increased coincidence rate (it will also produce an
increased rate of double events through scattering).
Fig.~\ref{fig:ddsolid}b presents the distribution $D(\theta)$ for the
double events, where 2 values of $\theta$ are used (one for each
detector involved in the double). $D(\theta)$ shows the same bias as
the $S(\theta)$ distribution.  Indeed, the $\theta$ angular variations
in the single and double count rates are highly correlated.  This can
be seen in Fig.~\ref{fig:ddsolid}c, which shows that although the
$\chi^2_\nu$ values for a constant fit to the variations of
$S(\theta)$ and $D(\theta)$ separately are both very high (giving
${\rm Prob.}<10^{-50}$ in both cases), the $\chi^2_\nu$ value for their
ratio, $R(\theta)=D(\theta)/S(\theta)$, is close to unity, and
acceptable for a constant fit (${\rm Prob.}=0.54$).

If the double count events were due entirely to coincidences of
independent single-count events we would expect that $D(\theta)\propto
S(\theta)$.  However, if a significant fraction of double-count events
are due to scattering, and polarization of gamma-ray photons produces
scatters toward a preferred direction in $\theta$ and $\theta+\pi$, as
suggested by the results of CB03, then we should observe a $D(\theta)$
that was significantly different than $S(\theta)$.  Our result here
conflicts with this, suggesting instead that the variability observed
by CB03 is not due to polarization-effected scattering, but due to
systematic uncertainty in the calculation of their ``null'' lightcurve by
MC simulation.

As a result, the coincidence rate for any given pair of detectors will
increase when they are both on the preferred side of the spacecraft,
and this may allow the derived ``scatter angles'' for the coincidence
events to pick out a preferred direction on the sky.  In the next
section we describe and perform a polarization analysis which is
insensitive to contamination from angular structure in the coincidence
events, which is clearly the dominant effect in the observed
double-count event variability. 

\section{A Mass-Model Independent Polarization Analysis}
\label{sec:anal}

We describe in this and the following sections a method for detecting
polarization which does not rely upon the mass model of \rhessi.

The observed countrate in a detector $i$ is: 

$$
N_i(t) = f(t) A_i(t)
$$

\noindent where $f(t)$ is the time dependent flux (\cgsflux) and
$A_i(t)$ is the detector response (counts erg$^{-1}$), which we allow
to be time-dependent due to the rotation of \rhessi\ with respect to
the sky.  We assume that the probability that a count: (1) is a
single-scattering event in detector $i$; (2) escapes detector $i$; and
(3) is then partially or fully absorbed in a second detector $j$, is a
time-independent factor $B_{i,j}$, dependent only on the materials and
geometry of the \rhessi\ spectrometer.

The number of double-count events due to unpolarized photons between
detectors $i$ and $j$ at a time $t$ when the centers of those two detectors produce
a line which has a position angle $\theta_{i,j}$ on the sky (relative to the
north celestial pole) can be found:

\begin{equation}
\label{eq:one}
N_{i,j}(t) = N_i(t) B_{i,j}  +  N_j(t) B_{j,i} 
\end{equation}

\noindent If the beam is polarized, this produces a polarization
pattern $I_p(\theta_{i,j})$ where $\theta_{i,j}(t)= \omega t +
\phi_{i,j}$ is now the angle the detectors make on the sky as a
function of time, and $\omega$ is the rotational frequency of the
\rhessi\  detector plane, and $\phi_{i,j}$ is the angle the detector pair
$(i,j)$ makes on the sky at time $t=0$.\footnote{We do not correct the
polarization angle measured at the detector, which is projected
relative to the North Celestial Pole, to the angle relative to the
north from the GRB position in the sky. The difference varies as a
function of polarization angle by up to 3\degree\ systematically.
This is $<$20\% of our 15\degree\ bins, and will not affect or
conclusions.}

\begin{equation}
N_{i,j}(\theta_{i,j}(t)) = N_i(\theta_{i,j}(t)) \, I_p(\theta_{i,j}) \, B_{i,j}  +
N_j(\theta_{i,j}(t))\, I_p(\theta_{j,i}) \, B_{j,i} 
\end{equation}

\noindent where we use the polarization intensity pattern: 
\begin{equation}
\label{eq:ip}
I_p(\theta) = \frac{1 + p\, \cos(2(\theta-\theta_p))}{1+p}
\end{equation}

\noindent (This differs by $\pi/2$ from a different, oft-used
definition).  Now, $I(\theta_{i,j})=I(\pi + \theta_{i,j})=I(\theta_{j,i})$; and
$B_{i,j}=B_{j,i}$ for the identical detector geometries which are used
by \rhessi.  Thus:

\begin{equation}
N_{i,j}(\theta_{i,j}) = [N_i(\theta_{i,j}(t)) + N_j(\theta_{i,j}(t))] B_{i,j} I_p(\theta_{i,j})
\end{equation}

\noindent Each side of this equation can be summed over all detector
combinations, keeping data binned as a function of $\theta$; then 

\begin{equation}
\sum_{i,j>i}^{\theta<\theta_{i,j}<\theta+\Delta_\theta}N_{i,j}(\theta_{i,j})
= I_p(\theta_{i,j})\, \sum_{i,j>i}^{\theta<\theta_{i,j}<\theta+\Delta_\theta}
[N_i(\theta_{i,j}(t)) + N_j(\theta_{i,j}(t))] \, B_{i,j} 
\end{equation}

\noindent which we can then re-write as a ratio, which is a function
of sky-angle only: 

\begin{eqnarray}
\label{eq:R}
R(\theta) & =  &  \frac{\sum_{i,j>i}^{\theta<\theta_{i,j}<\theta+\Delta_\theta}N_{i,j}(\theta_{i,j})}{ \sum_{i,j>i}^{\theta<\theta_{i,j}<\theta+\Delta_\theta}
(N_i(\theta_{i,j}) + N_j(\theta_{i,j}))} \\
& = & C \, \frac{1 + p \cos(2(\theta-\theta_p))}{1 + p}
\end{eqnarray}

If $p=0$ (unpolarized photons) then $R(\theta)$ is a constant, with no
dependence on $\theta$.  However, if $p\neq0$, then the observed
function $R(\theta)$ will have an angular dependence, as
$I_p(\theta)$. 

The only contribution to the uncertainty in $R(\theta)$ is photon
counting statistics.  More specifically, $R(\theta)$ is independent of
the mass model of the spacecraft, and the uncertainty of detection of
polarization using $R(\theta)$ does not depend on systematic
uncertainties in the space-craft mass model as it does in CB03.
However, it should be noted that correcting a detected signal to find
the intrinsic polarization magnitude (that is, turning the detection
into a quantitative measurement) would still require using the mass-model.

In practice, the function $R(\theta)$ is constructed from a discrete
sum of double-count events and a discrete sum of single-event counts,
as we describe in the following subsection.

Here, we note an additional correction.  As discussed in
\S\ref{sec:simple}, the total number of single events observed during
the GRB $I_i$ varies with detector $i$, which implies different
sensitivities to events.  If we take the relative sensitivity 
$f_i=I_i/\langle I_i \rangle$ to be independent of the detector
geometry (which are assumed identical), then Eq.~\ref{eq:one} becomes:

\begin{equation}
\label{eq:oneb}
N_{i,j}(t) = N_i(t)\, f_j B_{i,j}  +  N_j(t) f_i\, B_{j,i} 
\end{equation}

\noindent Here and above, $B_{i,j}$ is a relative solid angle
subtended by detector $j$ from detector $i$.  This correction affects
Eq.~\ref{eq:R}, which becomes: 

\begin{equation}
\label{eq:R2}
R(\theta)  =    \frac{\sum_{i,j>i}^{\theta<\theta_{i,j}<\theta+\Delta_\theta}N_{i,j}(\theta_{i,j})}{ \sum_{i,j>i}^{\theta<\theta_{i,j}<\theta+\Delta_\theta}
(N_i(\theta_{i,j})\, f_j + N_j(\theta_{i,j})\, f_i)} \\
\end{equation}

\label{sec:fdefinition}

\subsection{Data Analysis for \grb} 
\label{sec:dataanal}
We downloaded data from the public \rhessi\
archive\footnote{http://hesperia.gsfc.nasa.gov}, which also makes
available data analysis software we used -- the
Solar SoftWare (SSW) system written in IDL.  We
extract counts from photons with energy between 0.15 and 2 MeV during
the observation of the GRB roughly corresponding in time to those
analysed by CB03 (see Fig.~\ref{fig:lc}).  We find 85387 such counts
in the 5.0 sec interval we analyze, chosen to coincide with that
analysed by CB03.

The background number of counts, measured during an earlier
observational period, is 8636\ppm42 counts during the 5.0 second
period.

We find 8230 double-count events (\S\ref{sec:dcounts}).  To
perform the analysis, we use a discrete summation for the lightcurve.
To construct the 36 independent double-event light-curves
$N_{i,j}(\theta_{i,j})$, we take:

\begin{equation}
\label{eq:theta}
\theta_{i, j}(t_k)= (\omega t_k + \phi_{i,j}) \, {\rm mod}\,  (\pi)
\end{equation}

\noindent where $t_k$ is the time of the earliest time-stamp of each
double-count event.  The values of $\phi_{i,j}$ are relative angles of
the line joining the centers of detectors $i$ and $j$, and the centers
of detectors 1 and 2, except for $\phi_{1,2}$, which we set equal to
the angle on the sky at time $t=0$ clockwise of N made by the
center-line joining detectors 1 and 2, and $\omega=2\pi/({\rm 4.09\,
s})$ (the rotational frequency of the spacecraft at the time of the
observation).

The angles $\phi_{i,j}$ were found using the coordinate positions for each
detector as listed in Table~\ref{tab:spex}.  We divide the 180
degree range into 12 equal non-overlapping segments, beginning with
$\theta_n=n\times15$\degree, with $n=[0,11]$.  We produce
the summed double-event lightcurve dependent upon sky angle
$D(\theta_n)$: 

$$
D(\theta_n)= \sum_{k}N_{i,j}(\theta_{i,j}(t_k))
$$

\noindent Where $N_{i,j}$ takes the value of 1, for a double-event
involving detectors $i$ and $j$ with time-stamp $t_k$.  The angles are
binned as $\theta_n<\theta_{i,j}<\theta_n+\delta_\theta$.

Thus, for each double-event at time $t_k$ involving detectors $i$ and
$j$, we calculate the angle on the sky from Eq.~\ref{eq:theta}, and
add a count to the appropriate $\theta$ bin.  This is shown in
Fig.~\ref{fig:ddsolid}e.

We then construct the denominator from the single-event light curve: 

$$
S(\theta_n) = \sum_k \sum_{i_k,j!=i_k}  f_j\, N_{i_k}(\theta_{i_k,j}(t_k))
$$

\noindent where $N_{i_k}$ takes the value of 1, $f_j$ is described in
\S\ref{sec:fdefinition}, and for every event $k$ in detector $i_k$,
we calculate the angle on the sky for every detector pair which
includes $i_k$\footnote{We note that using $f_i=1$ instead of the
relative apparent counts sensitivity, as if only the physical apparent
cross sections $B_{i,j}$ were important, does not
change our conclusions.}.  Thus, for each count $k$, detected at time
$t_k$ in detector $i_k$ we find the eight angles on the sky
$\theta_{i_k,j}$ produced by the $i_k$ and $j$th detector, and add one
count in each of the eight bins, including angle bins in which the
same count falls more than once due to co-aligned detector pairs (such
as detector pair 4 \& 7 and pair 4 \& 3).  This is shown in
Fig.~\ref{fig:ddsolid}d.

We produce $R(\theta_n)=D(\theta_n)/S(\theta_n)$, which is shown in
Fig.~\ref{fig:ddsolid}f.  We discuss this Figure in detail in
\S\ref{sec:results}, after first discussing the effect of the
Klein-Nishina cross section on the detection in \S\ref{sec:kn}.

The uncertainty in $R(\theta_n)$ is set by the photon counting
statistics of $D$ and $S$.  As a minor correction, it is the number of {\em
unique} counts in each bin of $S$ which contributes to the
fractional uncertainty of $S$.  Thus, if there are 40000 counts in a
particular angle bin of $S$, but only 30000 of those are unique,
then the contribution to the fractional uncertainty of $R(\theta)$ is
$1./\sqrt{30000}$.  Even so, the uncertainty is dominated by the
smaller number of double-event counts, which are in the range 610-770
in each bin.

\subsection{Effect of the Klein-Nishina Cross Section}
\label{sec:kn}

The GRB was offset from the pointing axis, which is perpendicular to
the plane of detectors, by 17.98\degree.  Assuming a thin detector
plane\footnote{In fact, the detectors are 8\,cm long, comparable to
their width and separation. We neglect this width throughout the
analysis.}, photons which first scatter in a detector on the side of
the detector plane toward the GRB can scatter through an angle (to the
second detector) of up to $\theta_{\rm KN}$=90\degree +18\degree =108\degree;
while photons which first scatter on the side of the detector plane
away from the GRB could scatter through an angle of up to
$\theta_{\rm KN}=$90-18=72\degree.

The difference in scattering geometry produces a varying scattering
cross section for gamma-rays in the 0.15-2.0 MeV energy range,
dependent on $\theta$ -- the angle between a line connecting the centers
of the two detectors and the projection of the direction of the GRB
from the optical axis in the detector plane -- through the
Klein-Nishina cross section: 

\begin{eqnarray}
\frac{d\sigma}{d\Omega}(\theta_{\rm KN}) \propto \left(\frac{k'}{k}\right)^2 \left(\frac{k}{k'} +
\frac{k'}{k} - \sin^2\theta_{\rm KN}\right) \\
\frac{k'}{k} = \frac{1}{1 + \frac{h\nu}{m_e c^2}(1 - \cos \theta_{\rm KN})}\\
\end{eqnarray}

\noindent This will produce directionally preferential scattering from
a non-polarized beam of gamma-rays in the detectors.  If the direction
of the GRB is $\overrightarrow{k}= \sin\phi \widehat{x} + \cos\phi
\widehat{z}$ (where $\phi$ is 18\degree ) the line which joins two
detectors as $\overrightarrow{r} = \cos\omega t \widehat{x} +
\sin\omega t \widehat{y}$, then $\gamma(t)=\omega t$ and $\theta_{\rm
KN}(t)=\arccos(\sin\phi \, \cos\omega t)$.

In Fig.~\ref{fig:kn}a, we plot the KN cross section as a function of
$\gamma$, for $\phi=$ 18\degree, and for four photon energies ranging
from 0.15-2.0 MeV, normalized to its maximum value.  The cross section
changes by 20\% for photons of energy 0.15 MeV, and by 35\% for
photons of energy 2.0 MeV as a detector pair rotates in the detector
plane.

A non-polarized gamma-ray beam will produce an angle-dependent
double-event count-rate in the \rhessi\ detectors proportional to the
sum of the cross section in two detectors separated by 180\degree :

\begin{equation}
I_{\rm KN}(\gamma)\propto \frac{d\sigma}{d\Omega}(\theta_{\rm KN})  +
\frac{d\sigma}{d\Omega}(\theta_{\rm KN} + \pi)
\end{equation}

\noindent We show $I_{\rm KN}(\gamma)$ in Fig.~\ref{fig:kn}b.  The KN
cross section results in an angle-dependent variation in total
intensity of 7-9\% across the 0.15-2.0 MeV band.  For comparison, we
include in this figure $I_p(\theta)$, for $p=0.045$.  To within the
our observational uncertainties, the KN cross-section can always be
represented by $I_p(\theta)$\footnote{While the $I_{\rm KN}(\theta)$ is
not identically parameterizable by $I_p(\theta)$ we find that
parameters to $I_p(\theta)$ can typically be found which duplicates
$I_{\rm KN}(\theta)$ to better than 1\% for $p<0.1$.}.  It is notable that
the KN cross-section produces an effect on the observed signal which
is nearly identical to that expected from an intrinsically polarized
beam.

The magnitude of ``apparent polarization'' (that is, the value of $p$
which parameterizes $I_{\rm KN}(\theta)$ using $I_p(\theta)$; see
Fig.~\ref{fig:kn}c) varies as a function of photon energy; it is
$p=5$\% (for photon energy 0.15 MeV), increasing to 5.1\% (near 0.30
MeV) and decreasing to 3.7\% a 2 MeV.

This effect is easy to correct for during data analysis, if one knows
the direction of the GRB relative to the detector pairs.

We calculated the fractional magnitude and direction of the KN effect
on the double-count scattering events, by averaging the magnitude of
the effect at the energy of each detected count during the GRB
(neglecting the 10\% of counts due to background) and finding the
average and standard deviation, which were $\langle p_{\rm KN}
\rangle=$0.049\ppm0.003. We can represent the effect of the KN cross
section on the intensity pattern produced by the GRB by:

$$
I_{\rm KN} = \frac{1 + 0.049\, \cos(2(\theta-\theta_{\rm GRB}))}{1.049}
$$

\noindent where $\theta_{\rm GRB}$ is the direction of the GRB
projected onto the detector plane ($\theta_{\rm
  GRB}=45$\degree\ E of N).  

In the present data, the magnitude of $p_{\rm KN}$ will be decreased
by a factor $I_{\rm 2, scattered}/I_{\rm 2, observed} (<0.046)$ (90\%
confidence). Thus, in the present data, $\langle p_{\rm KN}
\rangle<0.0022$, which, as we show in \S\ref{sec:results}, is well
below the detection limit.

Moreover, the magnitude of this signal is overestimated, as we have assumed
infinitely thin detectors.  The effect of finite sized detectors is to
diminish the magnitude of this effect, as can be immediately seen, for
example, in the limit of infinitely long detectors.  Moreover, the
need for the correction in the present work is absent; the maximum
magnitude of the effect is 0.049$\times$830\ppm150=41\ppm7 counts --
smaller than the Poisson noise of the total double-count events (which
is $\sqrt{8230}=91$ counts).  Thus, this effect will not be important
in the present analysis.

\subsection{Results}
\label{sec:results}

In Fig.~\ref{fig:ddsolid}f, we show $R(\theta)$ for the observation of
\grb.  First, $R(\theta)$ is consistent with being a constant (reduced
chi-squared value \chisqrnu=1.35 for 11 dof).
We therefore find no evidence of a signal which might be interpreted
as due to polarization in the \rhessi\ data for \grb.

To obtain a limit on a signal which might be interpreted as due to
polarization, we find a best-fit for $I_p(\theta)$ (\chisqrnu=1.36, 9
dof) with $p\leq0.041$, or $\leq$4.1\% (90\% confidence), finding a
(likely, local) minimum at $\theta_p=74$\ppm22\degree.  A similar
limit on $p_{\rm KN}$ is  obtained holding fixed\ $\theta_{GRB}=45$\degree,
($p_{\rm KN}<0.04$), which is consistent with our calculated
upper-limit for this effect.

\section{Duplication of CB03 Results}
\label{sec:bad}

We attempt to duplicate the results of CB03. In particular, we are
interested in duplicating their observed double-event lightcurve.
While the methods of producing the double-event lightcurve of CB03
have not been described as of this writing, there are only a few
variables which were not specified by CB03: (1) $\Delta T$; (2)
whether or not multiple-scattering events ("bunches") are retained;
and (3) the zero point for $\theta$.

We show the raw (double-event) counts and residual counts as a
function of $\theta$ (\itheta\ hereafter) obtained by CB03 in
Fig.~\ref{fig:cb03}a.  The residual counts lightcurve is the difference
between the raw counts and their MC estimation for the double-event
count rate ($M(\theta)$=\itheta$-$\intheta), assuming zero intrinsic
polarization, and using the GEANT mass model for radiative transfer
through \rhessi\
\footnote{It was this MC simulation which gave the estimate of
$\Pi=0.19\ppm0.04$ -- that of all photons which are detected as
double-count events, 19\ppm4\% have not been scattered previously in
spacecraft passive material, and thus have the polarization intact
during measurement.}.  We note that the average values of \itheta\ and
\intheta\ found by CB03 coincide within $<$0.1\%.

We obtain a good duplication of the CB03 double-event count curves,
using $\delta T$=8\, \bms\ (as compared with $\Delta T$=5\, \bms\ in our
previous analysis); retaining all multi-count events (``bunches'',
including triples and higher); and setting $\theta=0$ to coincide with
12.6\degree\ E of N in the Ecliptic (Fig.~\ref{fig:cb03}b, top panel).
Here, we find $N_2$=15540 counts, which is  close to the
$N_2$=14916 found by CB03.  We rescaled the \intheta\ calculated by
CB03 so that $\langle \intheta \rangle = \langle \itheta \rangle$.
The residual counts (\itheta$-$\intheta; Fig.~\ref{fig:cb03}b, bottom
panel) is similar to that obtained by CB03 (Fig.~\ref{fig:cb03}a,
bottom panel).

Using these same selections, we apply our $R(\theta)$ analysis.  If
the modulation seen in Fig.~\ref{fig:cb03}b (bottom panel) is due to
polarization, we expect to see the a sinusoidal modulation in
$R(\theta)$, too (see Eqns.~\ref{eq:R} \& \ref{eq:R2}).  Instead
(Fig.~\ref{fig:cb03}c), we see that $R(\theta)$ is scattered,
significantly different from a constant (\chisqrnu=2.49/11 dof;
prob.=0.4\%), but with no systematic trend approximating a sinusoidal
modulation. The scatter in $R(\theta)$ cannot be due to a constant
polarization signal as implied by CB03 (as we
presently show, the scatter is due to the ``bunches'').  We conclude
from this that the signal observed by CB03 is not due to polarization
in \grb.

We then dropped from the analysis all the ``bunches'' of counts
(multiplicities of 3 and greater).  We do this, because we see no
reason that polarized photons should be preferentially detected in
events which involve $\geq$2 scatters. Moreover, as we have shown
above, the bunches do not follow a Poisson distribution, and do not
follow a simplistic distribution expected from scattering; we
therefore interpret these ``bunches'' as a background signal.  There
were a total of 1272 such bunches, producing a total of 4592
double-count counts; after removal, there were 10948 double-count
events remaining.  We performed the analysis again: \itheta\ and
\intheta\ in Fig.~\ref{fig:cb03}d, top panel; with residual counts
\itheta$-$\intheta in the bottom panel.  While the modulation of the
residual counts is reduced, there remains a significant deviation (in
the range $\theta$=100-180\degree).

However, the corresponding $R(\theta)$ (Fig.~\ref{fig:cb03}e) is
consistent with a constant value (\chisqrnu=0.66, 11 dof, prob.=77\%).
Applying the identical analysis as above, we again obtain a limit of
$p<0.041$ (90\% confidence).  This means that polarization is not
detected (and that the scatter in $R(\theta)$ in Fig.~\ref{fig:cb03}c
is due to the presence of the ``bunches'' of counts).

When we duplicate the analysis of Sec.~\ref{sec:simple} on the data in
Fig.~\ref{fig:cb03}b and c, we obtain a fraction of counts which may
be due to double-count scattering events of $f$=9\ppm2\%
(\chisqrnu=6.0, 6 dof; Prob.=3\tee{-6}) and 8\ppm3\% (\chisqrnu=13.9,
6 dof; Prob.=0.03), respectively.  The fact that data which includes
the ``bunches'' is not well-modeled by our scattering model is
consistent with the conclusion that these events are not related to
scattering.

To what can the modulation reported in CB03 be due?  Since the lack of
modulation in $R(\theta)$ implies that $\itheta\propto S(\theta)$, yet
$\itheta-\intheta$ is not constant in $\theta$, the modulation must be
in \intheta, and not in \itheta.  We attribute, therefore, the
modulation observed by CB03 to neglected systematic uncertainty in
\intheta, which we explore in \S~\ref{sec:systematic}.

\section{The Intrinsic Polarization of \grb}

Our parameter $p$ is not the intrinsic fractional polarization $\Pi$
of \grb.  The relationship of our parameter $p$ to $\Pi$ is:

\begin{equation}
p = \mu \Pi \frac{S}{S+B}
\end{equation}

\noindent where $\Pi$ is the intrinsic fractional polarization of
\grb; $\mu$ is the fraction of \rhessi-detected double-count photons
which have not been previously scattered within the detector; $S$ is
the number of double-count events due to scattering, and $B$ is the
number of background double-count events (that is, not due to
scattering).

If our parameter $p$ is consistent with zero, then the intrinsic
polarization $\Pi$ is also consistent with zero; thus, regardless of
the magnitude of (and uncertainties in) $\mu$, $S$ and $B$, our
upper-limit of $p<0.041$ implies a non-detection of polarization in
the \rhessi\ data of \grb.  This is in conflict with the claim of
detection described by CB03.  

We now describe the origin of the discrepancy between these two
results.

Based on our limit $p<0.041$, $S=830\ppm150$ counts, and $S+B=8230$
counts, we find the 90\% confidence upper-limit for intrinsic
polarization in \grb\ is $\Pi<0.41/\mu$.

CB03 finds $\mu=0.19\ppm0.04$ from their Monte Carlo mass-model
simulations, which we adopt.  Relying upon this value, $\Pi<2.14$
(fractional polarization, or $<$214\%) -- that is, an upper-limit which
permits up to 100\% intrinsic polarization for \grb, implying that the
observation is not sensitive to intrinsic polarization in \grb.  

The lack of sensitivity is due to the lack of double-count events from
scattering; the number we found (830\ppm150 counts) is smaller than
and in conflict with the number found by CB03 (9840\ppm96).

If we had found the fraction $f$ of total double-count events from 
scattering to be that found by CB03 ($f=$66\ppm1\%; see
Table~\ref{tab:bg}), we would would instead conclude that $\Pi<0.46$
(90\% confidence), compared with the $\Pi=0.80\ppm0.20$ (CB03).
However, our analysis (\S\ref{sec:dcounts}) finds that $f$=11\ppm3\%,
which produces instead a limit on the intrinsic polarization of \grb\
well above 100\%.  

Our conclusion is that the \rhessi\ dataset is not capable of
detecting polarization in \grb, due to the low signal-to-noise of the
dataset (that is, a relatively small number of double-count scattering
events compared to that estimated by CB03).  

\section{Systematic Uncertainty in CB03}
\label{sec:systematic}

In our analysis, the uncertainty in polarization {\em detection} is
due only to Poisson noise (in $R(\theta)$); in the analysis of CB03,
the uncertainty in polarization {\em detection}
($M(\theta)=D(\theta)-D_{\rm null}(\theta)$) is the sum of Poisson
noise and the systematic uncertainty in $D_{\rm null}(\theta)$.  CB03
neglect the systematic uncertainty in $D_{\rm null}(\theta)$.  Here,
we show that this systematic uncertainty is not negligible. 

The observable modulation lightcurve is:

$$ 
M(\theta)=D(\theta) - D_{\rm null}(\theta).
$$

The systematic uncertainty in $M(\theta)$ due to radiative transfer
through the spacecraft mass model appears  in $D_{\rm null}(\theta)$.
In a simple  formulation, one can represent this as:

$$
D_{\rm null}(\theta)=\langle D_{\rm null} (\theta) \rangle \,
\frac{\mu(\theta)}{\langle \mu \rangle}
$$

The function $\mu(\theta)$ and its uncertainty was not given by CB03.
To be able to neglect the systematic uncertainty in $\mu$, its
magnitude would need to be \approxlt\ half the Poisson uncertainty --
$\approxlt$20 counts, or 1.6\% per angular bin, given the $\langle
D(\theta) \rangle=1250$ counts/bin.

We can estimate the uncertainty in $\mu(\theta)$ in the analysis of
CB03 from the uncertainty in its mean value ($\langle \mu \rangle =
\mu=0.19\ppm0.04$), which has a 25\% uncertainty.  If we take the
uncertainty in $D_{\rm null}(\theta)$ to be the same fractional value
as the uncertainty in the average $\mu$ (25\% per angular bin), we
find approximately $\sigma_{\rm systematic}\approx0.25\times1250 ({\rm
counts/bin}) \approx$310 counts/bin, much larger than the 20 counts
limit, and which should appear directly in their difference lightcurve
(their Fig.2, or our Fig.~\ref{fig:cb03}a)\footnote {This is a
conservative estimate, as we have assumed the fractional systematic
uncertainty in the 12 individual $\theta$ bins is equal to that of the
average value of $\mu$; the fractional uncertainty in the individual
bins should be larger than that of the average value, by a factor of
$\sqrt{12}$ if they are Gaussian distributed.}.  However, their analysis shows only the Poisson uncertainty
($\sigma_{\rm Poisson}\sim$40 counts).  Moreover, the neglected
$\sigma_{\rm systematic}=310$ counts/bin is greater than the reported
modulation magnitude ($\sim$120 counts/bin); thus, we see no reason to
neglect it.

We conclude that the analysis for detection of polarization by CB03
incorrectly neglected the dominant uncertainty -- the systematic
uncertainty $D_{\rm null}(\theta)$, which must be added in quadrature
with the Poisson uncertainty.  This uncertainty, from the above
approximate analysis, is greater than the reported signal interpreted
as due to polarization.  

\section{Discussion and Conclusions}
\label{sec:con}

In selecting data to analyze, we found 8230 double-count events
(including signal and irreducible background). This is well below the
14916 double-count events found by CB03 (Table~\ref{tab:bg}).  The
discrepancy appears to be due to two selections employed by CB03: (1)
an unjustifiably wide time window for ``simultaneous'' events ($\Delta
T=8 \bms$, whereas we see only $\Delta T=5 \bms$ as being justified)
and (2) the inclusion of multi-count events (``bunches''), in which
$>$2 counts are detected within $\Delta T$.

We described an analysis to detect polarization in \rhessi\ data.  We
apply this analysis to the data for \grb, and find no evidence for a
signal which might be interpreted as due to polarization.  The
magnitude of polarization-like modulation in the lightcurve is
$p<$0.041 (90\% confidence).  This corresponds to an upper-limit on
the intrinsic polarization of \grb\ of $\Pi<$214\%; that is, we find that
the analysis is insensitive to polarization at any level in \grb.   

The discrepancy between our derived upper-limit and the claimed
detection by CB03 is due to two effects.  First, during {\em
detection}, CB03 neglected the systematic uncertainty in their null
lightcurve, due to the Monte Carlo simulation of radiative transfer
through the \rhessi\ mass model; the magnitude of this uncertainty is
consistent with the magnitude of the observed modulation.  In
comparison, this systematic uncertainty does not play a role in the
present $R(\theta)$ analysis, since the mass model is unnecessary.  We
are instead limited by Poisson noise.  Thus, we conclude that the
modulation interpreted as due to polarization by CB03 is instead due
to systematic uncertainty in their null lightcurve.

Second, during {\em correction} of the detection of polarization
signals into the magnitude of intrinsic polarization in \grb\ we find
11\ppm3\% of all detected double-count events in our selection are due
to scattering events.  This results in a low signal-to-noise ratio,
and a resulting high upper-limit on $\Pi$, such that the observation
does not constrain the intrinsic polarization of \grb.  We justified
our data selections and demonstrate that the signal should be clean of
irreducible background.  In contrast, CB03 estimated that 66\ppm1\% of
their detected double-count events were due to scattering events,
while not describing or justifying their selection
criteria. Specifically, our duplication of the analysis of CB03 found
an unjustifiably larger ``simultaneous'' window ($\Delta T$=8\, \bms,
vs. $\Delta T$=5\, \bms\ we used) and included the multi-count events
(``bunches'') -- both of which increase the background.  

In our analysis, the limiting sensitivity is the highest theoretically
possible -- set by photon counting statistics.  We therefore believe
our analysis is more robust, and we conclude that: (1) the \rhessi\
observation of \grb\ is not sensitive to polarization, and (2) there
is otherwise no evidence for polarization in the data.

In conclusion, we find that there is no existing constraint on the
intrinsic polarization in the gamma-ray flux of \grb\ -- or, for that
matter, any gamma-ray burst.  It seems unlikely that a constraint will
emerge from further \rhessi\ observations.  \grb\ was extremely bright
(brighter than any GRB observed during 8 years of full-sky coverage
with ULYSSES; \citenp{atteia99}) and detector deadtime already played
a role in decreasing the signal-to-noise ratio at the peak of the
burst, meaning an even brighter burst will only moderately improve the
signal-to-noise ratio; also, \grb\ was located within 5\% of the sky
closest to the solar limb, as it must to take advantage of the
polarization detection approach used herein (distinctly different from
that used to measure polarization in solar flares
\citenp{mcconnell02}), further decreasing the likelihood of a useful
constraint on GRB polarization with \rhessi.  Such a measurement must
therefore wait for more sensitive instrumentation.

\acknowledgements

We thank David Smith (UC Santa Cruz) for discussions regarding the
\rhessi\ mass model, and for useful conversations about the \rhessi\
spectrometer detectors, contributions to double-count event
backgrounds and data analysis; Wayne Coburn and Steve Boggs for
detailed conversations about their approach and conclusions; and
Andrew MacFadyen for comments on the manuscript.  We are grateful to
Gordon Hurford for detailed description and software to produce the
roll-angle solution for \rhessi\ at the time of the gamma-ray burst.
We acknowledge useful comments by the anonymous referee which
encouraged us to attempt to reproduce the results of CB03.

\clearpage

\figcaption{ \label{fig:spex} Diagram of \rhessi\ spectrometer
detector layout.  Each circle represents the 7.1\,cm diameter detector
unit. The numbering of the detectors corresponds to that in general
use, and these are shown as from the solar perspective.  The sense of
rotation shown is the rotation of the spacecraft; the axis of rotation
is not that shown, but is instead close to the coordinate origin, (0,0).  }

\figcaption{ \label{fig:diffdt} {\bf (a)} Histogram of time-stamp
  separation of consecutive counts in \grb, in units of $2^{-20}$ s
  (\bms). {\bf (b)} Same as (a), except consecutive counts in the same
  detector have been replaced by a single count at the earlier
  timestamp.  The peak at non-zero value of $\Delta T$ indicates that
  simultaneous counts can have time-stamps which are not equal.  We
  thus adopt the convention that counts with timestamps which differ
  by $\leq5$\, \bms\ are ``simultaneous''.  }

\figcaption{ \label{fig:lc} The 0.15-2.0 MeV lightcurve of GRB021206,
  as observed with \rhessi, beginning at 2002 Dec 6 22:49:10 UT.  The
  two vertical lines delineate the 5.0\,s period used in the present
  analysis, which were chosen to match the analysed data in CB03.}

\figcaption{\label{fig:derekfit}%
  Relative frequency of double-count events, by detector.  Top
  panel: Data points, with error bars, give the number of double-count
  events associated with each detector and their Poissonian
  uncertainties.  The solid histogram represents the best-fit scattering +
  coincidence model to these data, which has 11\ppm3\% scattering.
  Dotted histograms show the contributions of scattering (lower
  histogram) and coincidence (upper histogram), respectively.  Bottom
  panel: Residuals of the data compared to the model, scaled to the
  Poisson uncertainties.  All double-events that interact with
  detector 2 have been excluded from this analysis; see text for
  details.}

\figcaption{\label{fig:ddsolid}%
In these figures, $\theta=0 $ and $\theta_n$=0 correspond to celestial
north, with increasing value rotating east of north about the detector
origin (clockwise, facing the Sun; cf. Fig 1).  The position of \grb\
is marked in each panel by a vertical broken line.  {\bf (a)} Single
event countrate $S$ as a function $\theta$, where $\theta$ is the
angle between celestial north and the center of the detector in which
the count is observed, in the range 0-360\degree. The countrate is
inconsistent with being constant in $\theta$, with the \chisqrnu\
shown, and a probability of observing such variability of
$<$\ee{-100}.  {\bf (b)} Double-count event countrate $D$ as a
function of $\theta$; here, we place 1 count for each of the two
involved in the event, with each being placed in the appropriate
$\theta$ bin.  The strong variability is inconsistent with being
constant in $\theta$, with the given \chisqrnu, and a probability of
of observing such variability of $<$\ee{-50}.  {\bf (c)} The ratio
$S/D$ is consistent with being constant (prob=0.54).  {\bf (d)} The
single-count event rate $S$ in our polarization analysis, as a
function of sky angle $\theta_n$ in the range 0-180\degree\ (see
\S\ref{sec:anal}).  The countrate is inconsistent with being constant
in $\theta_n$.  {\bf (e)}The double-count event rate $D$ in our
polarization analysis, as a function of sky angle $\theta_n$, which
has values in the range 0-180\degree\ (see \S\ref{sec:anal}).  This
figure is directly comparable to the double-count event rate of CB03,
without having subtracted the Monte Carlo ``null'' lightcurve.  As in
CB03, we also find significant variability as a function of
$\theta_n$.  {\bf (f)} The ratio $R=D(\theta_n)/S(\theta_n)$ vs. sky
angle $\theta_n$.  Polarized photons would scatter through a preferred
angle on the sky, producing more double-count events in that direction
relative to single count events than in the perpendicular direction,
producing variation in $R(\theta_n)$. However, $R$ is observationally
consistent with being constant in $\theta$; thus, we find no evidence
of polarization.  }

\figcaption{ \label{fig:kn} {\bf Panel a:} The relative Klein-Nishina
cross-section for a pair of detectors as a function of angle between
the initial photon direction and a line between the two detectors, at
four different photon energies (listed top to bottom). {\bf Panel b:}
The relative cross section for a double-scatter event for a pair of
detectors due to the Klein-Nishina cross section, as a function of
angle between the detectors and the direction of the GRB projected in
the plane of the detectors, assuming the GRB is offset by 18\degree\
from the axis of the plane occupied by the detectors (zero offset
would produce zero modulation).  At 18\degree, the magnitude of
modulation is between 7\% (for 0.15 MeV photons) and 9.5\% (for 2.0
MeV photons). The solid lines are for the same photon energies as
given in panel a, with larger energies producing greater variation.
The dotted line is the $I(\theta)$ pattern {\em only} for a polarized
photon beam, with $p=0.045$ and $\theta_p=0$. {\bf Panel c:} The
inferred fractional polarization due only to the K-N cross-section, as
a function of photon energy. }
 
\figcaption{\label{fig:cb03} {\bf a)} Data copied from Fig.2 of
CB03. (Top Panel) Crosses are the double-count event lightcurve
\itheta\ as a function of angle on the sky $\theta$.  Solid squares
are the MC-simulated null-lightcurve $\intheta$.  Note the data is
repeated at 180\degree .  (Bottom Panel) Residual counts,
$\itheta-\intheta$, exhibiting the modulation in $\theta$ interpreted
by CB03 as due to polarization in \grb. The solid-line is their
best-fit of a polarization signal (also in panels b and d).  {\bf b)}
(Top Panel) Our duplication of the results of CB03, finding data
selections which reproduce their \itheta\ (see text), with symbols
having the same meaning as in panel a. The solid points, however, are
the \intheta\ values calculated by CB03, renormalized (see text).
(Bottom Panel) The residual counts curve duplicates well that of CB03.
{\bf c)} The ratio $R(\theta)=\itheta/S(\theta)$, using our duplicate
data selection.  The sinusoidal modulation in $\itheta-\intheta$ is
not present in $R(\theta)$; instead, the ratio shows significant
random scatter. {\bf d)} The same analysis as in panel b, but with the
multi-count events removed from the dataset.  The modulation in
\itheta-\intheta is reduced, but still present.  {\bf e)} The ratio
$R(\theta)$ using the same data set as panel d.  The scatter in panel
(c) has disappeared (due to the multi-count events removed), and no
modulation is present.  This implies that $\itheta\propto S(\theta)$
and that the modulation in $\itheta-\intheta$ seen by CB03 (panel a),
and duplicated here in panels b and c, is not due to
polarization-related modulation in $\itheta$, but in $\intheta$ (the
null lightcurve).  }

\clearpage
\pagestyle{empty}
\begin{figure}[htb]
  \PSbox{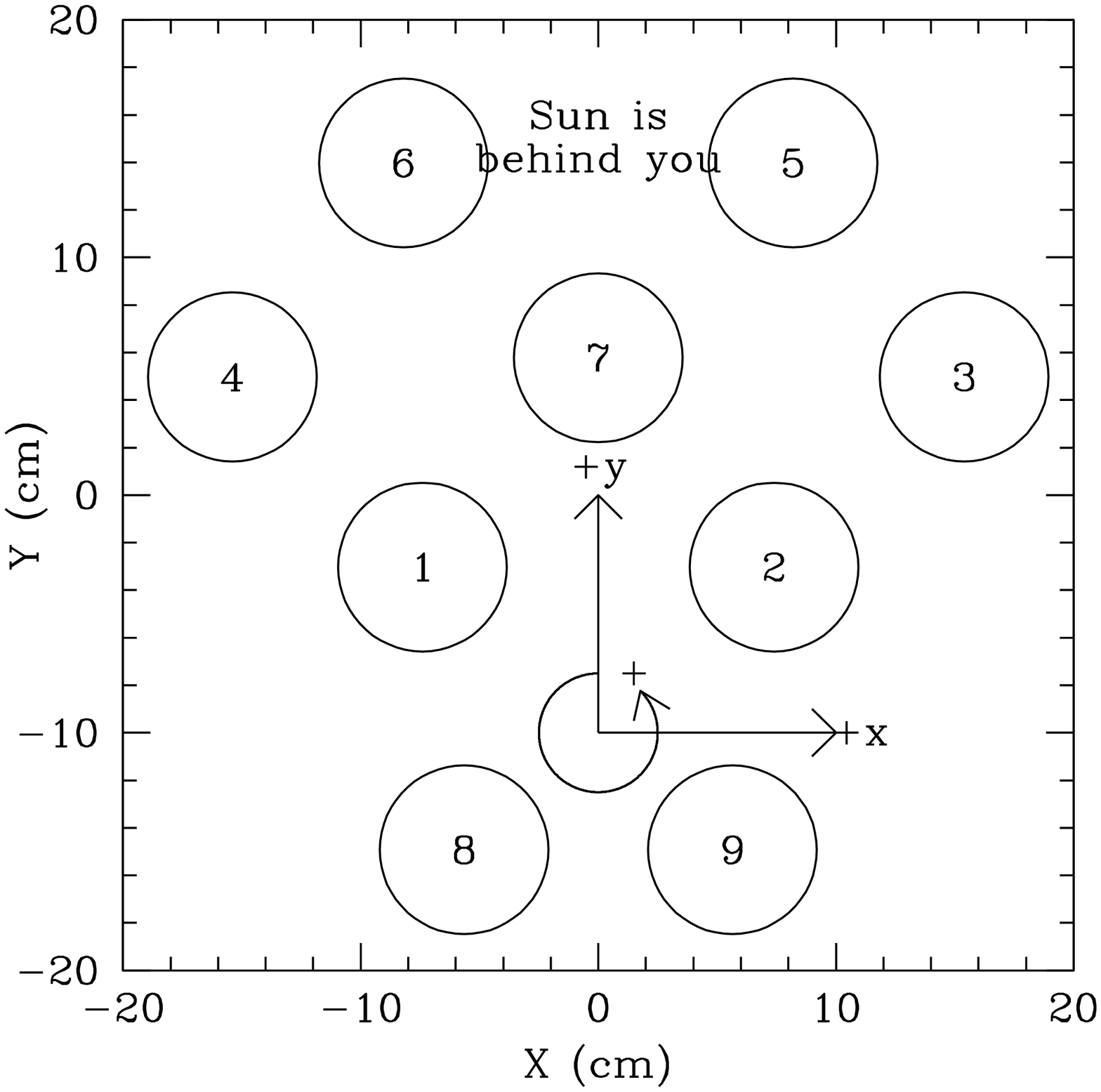 hoffset=-80 voffset=-80}{14.7cm}{21.5cm}
\FigNum{\ref{fig:spex}}
\end{figure}

\clearpage
\pagestyle{empty}
\begin{figure}[htb]
\PSbox{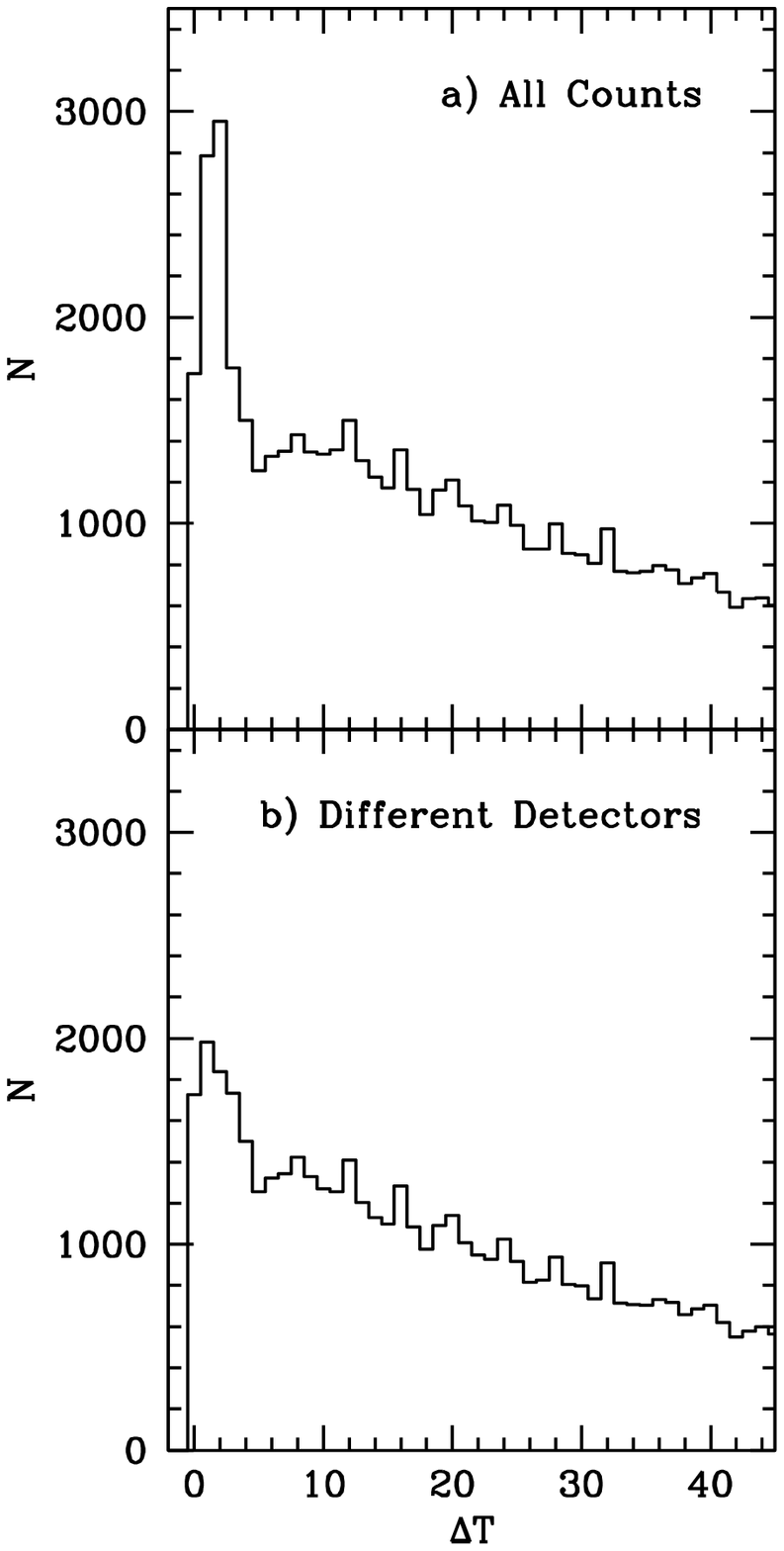 hoffset=-80 voffset=-80}{14.7cm}{21.5cm}
\FigNum{\ref{fig:diffdt}}
\end{figure}

\clearpage
\pagestyle{empty}
\begin{figure}[htb]
\PSbox{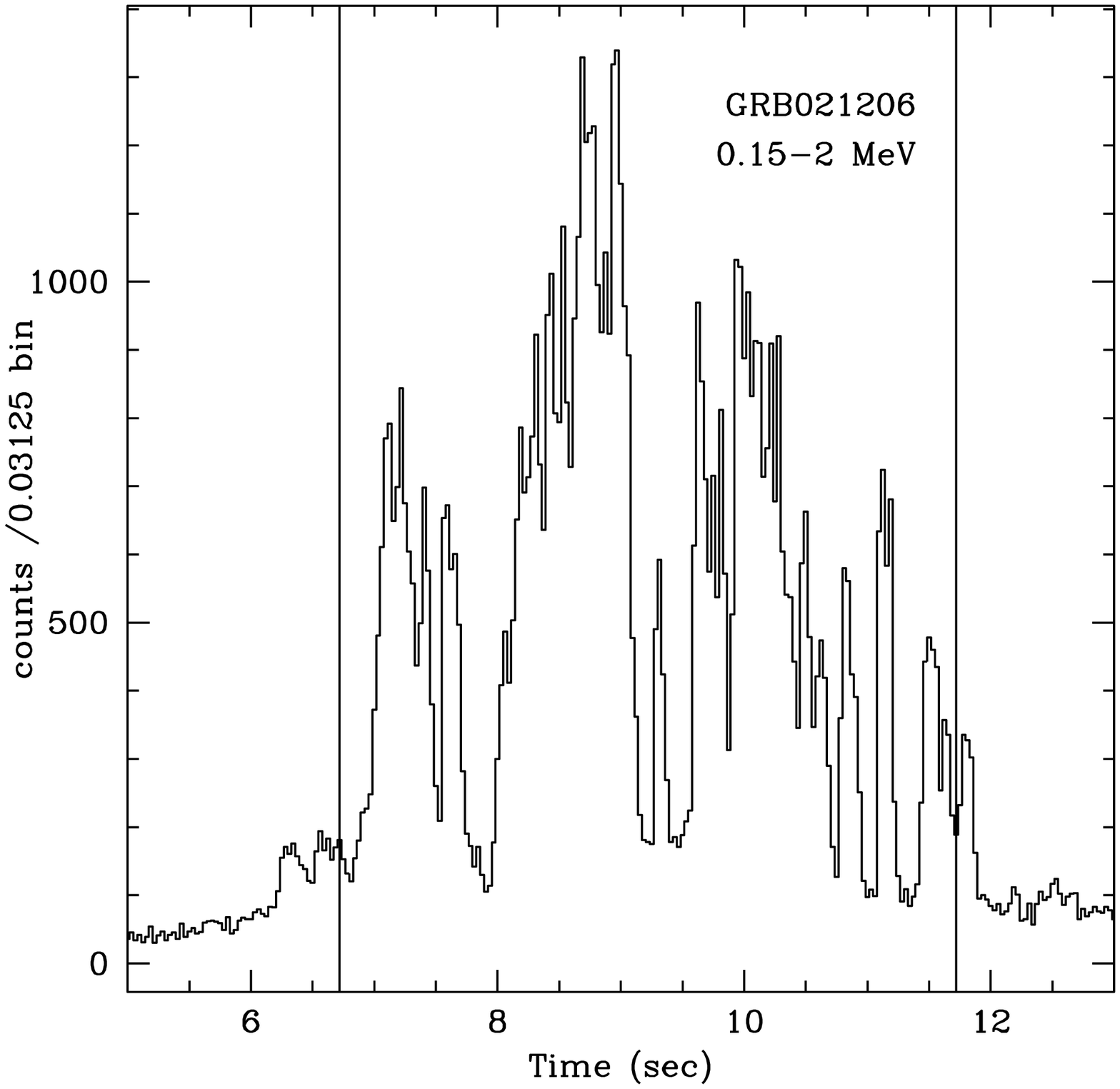  hoffset=-80 voffset=-80}{14.7cm}{21.5cm}
\FigNum{\ref{fig:lc}}
\end{figure}

\clearpage
\pagestyle{empty}
\begin{figure}[htb]
\PSbox{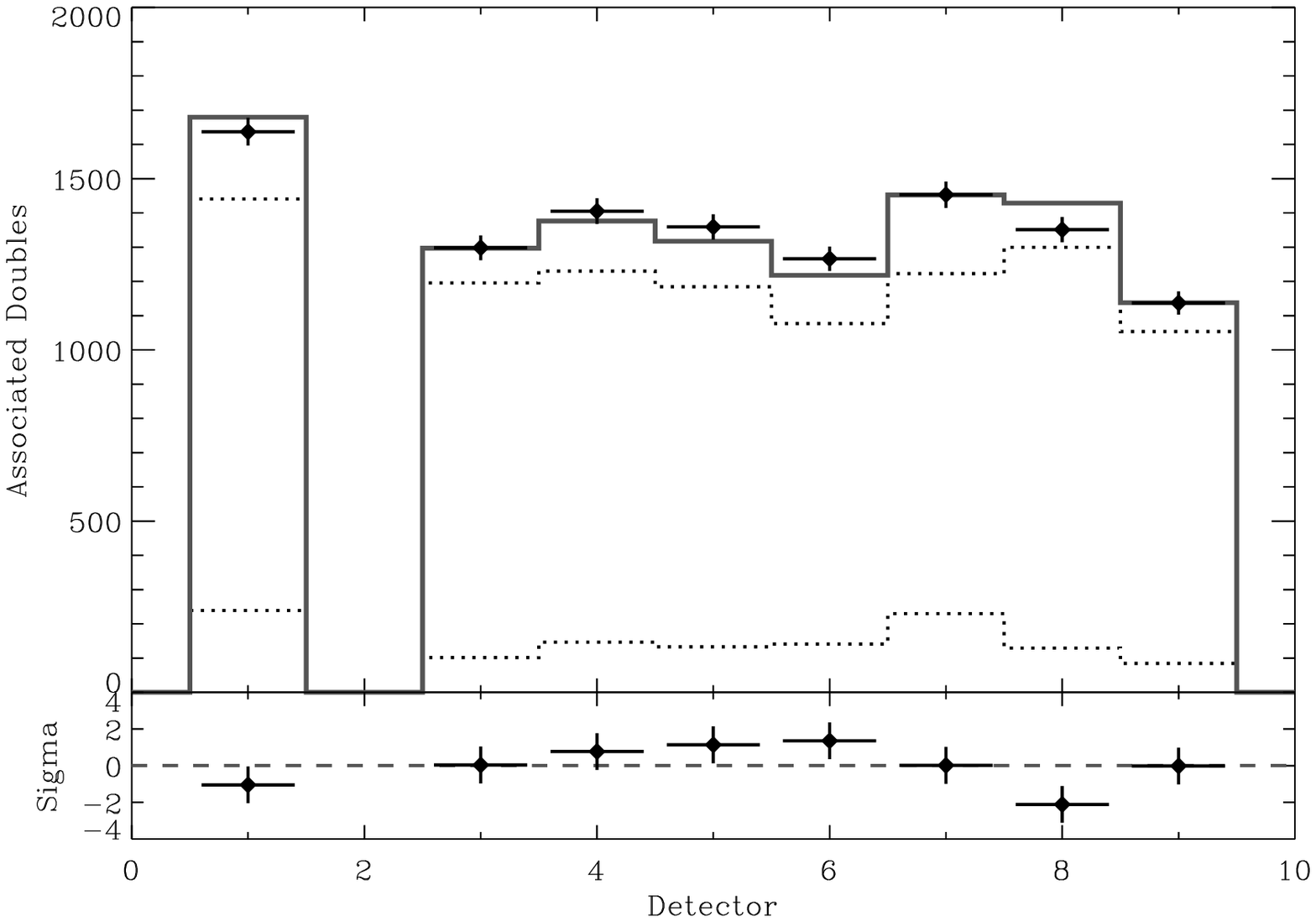 hoffset=-80 voffset=-80}{14.7cm}{21.5cm}
\FigNum{\ref{fig:derekfit}}
\end{figure}

\clearpage
\pagestyle{empty}
\begin{figure}[htb]
\PSbox{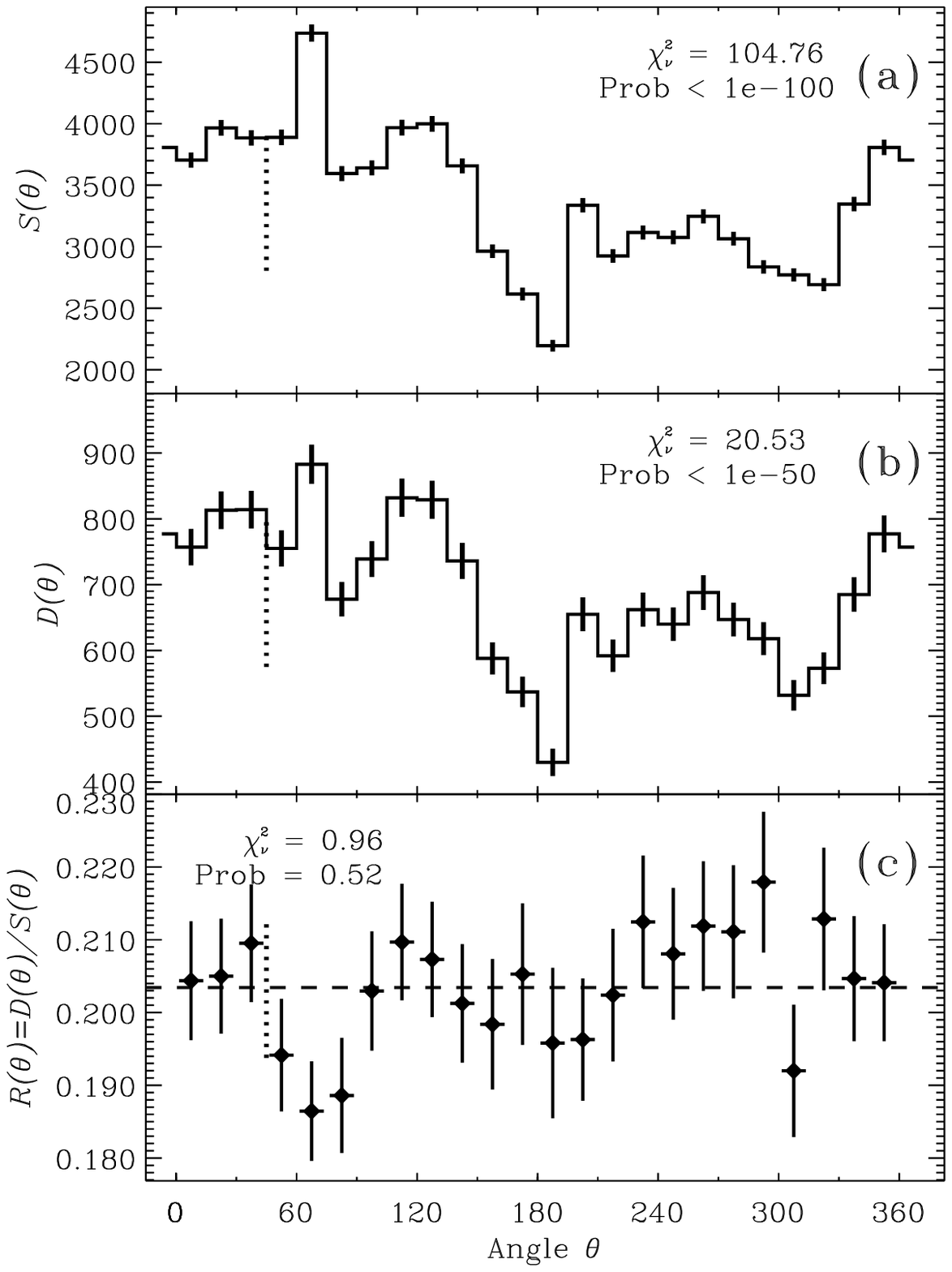 hoffset=-80 voffset=-80}{14.7cm}{21.5cm}
\FigNum{\ref{fig:ddsolid}a-c}
\end{figure}

\clearpage
\pagestyle{empty}
\begin{figure}[htb]
\PSbox{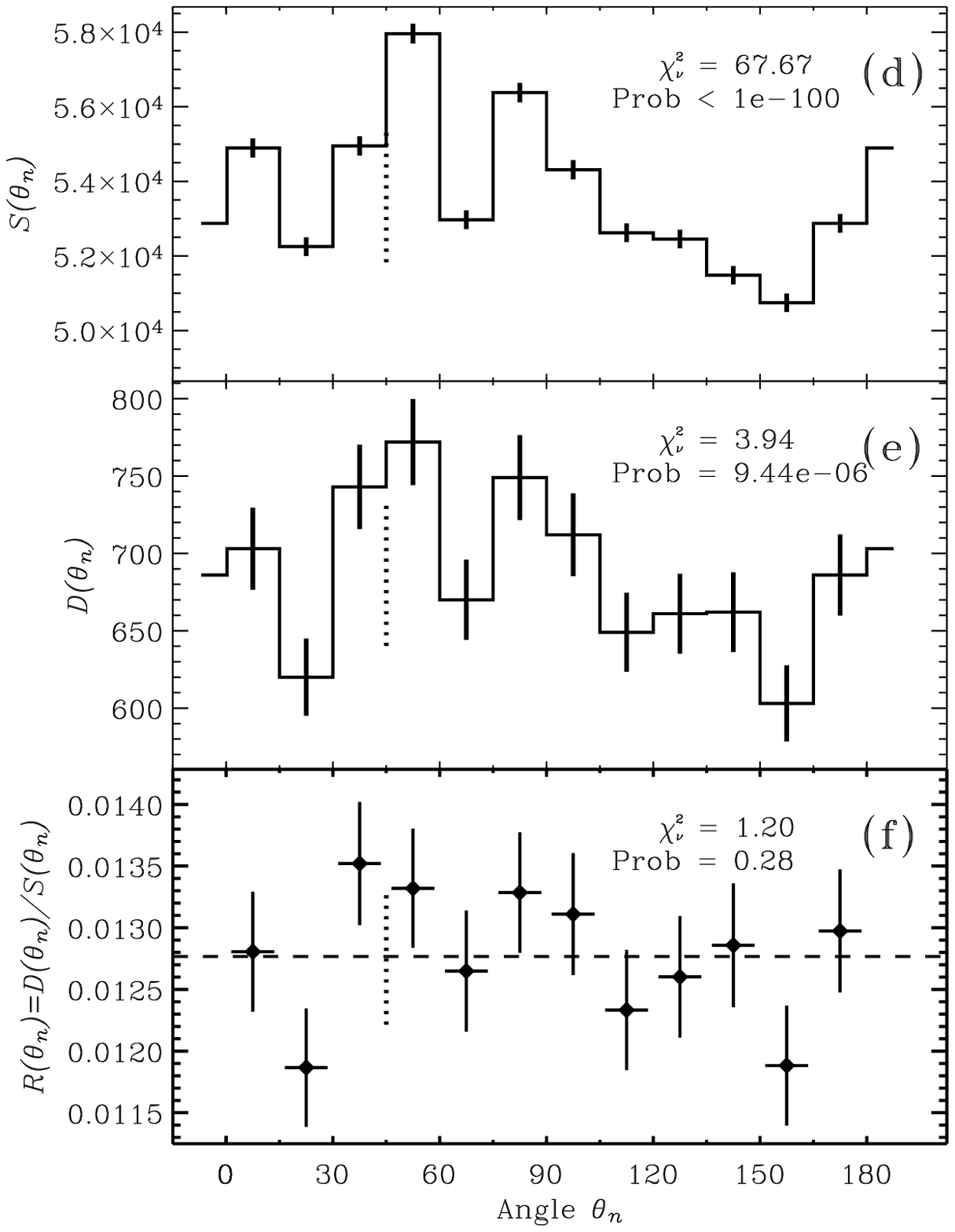  hoffset=-80 voffset=-80}{14.7cm}{21.5cm}
\FigNum{\ref{fig:ddsolid}d-f}
\end{figure}

\clearpage
\pagestyle{empty}
\begin{figure}[htb]
\PSbox{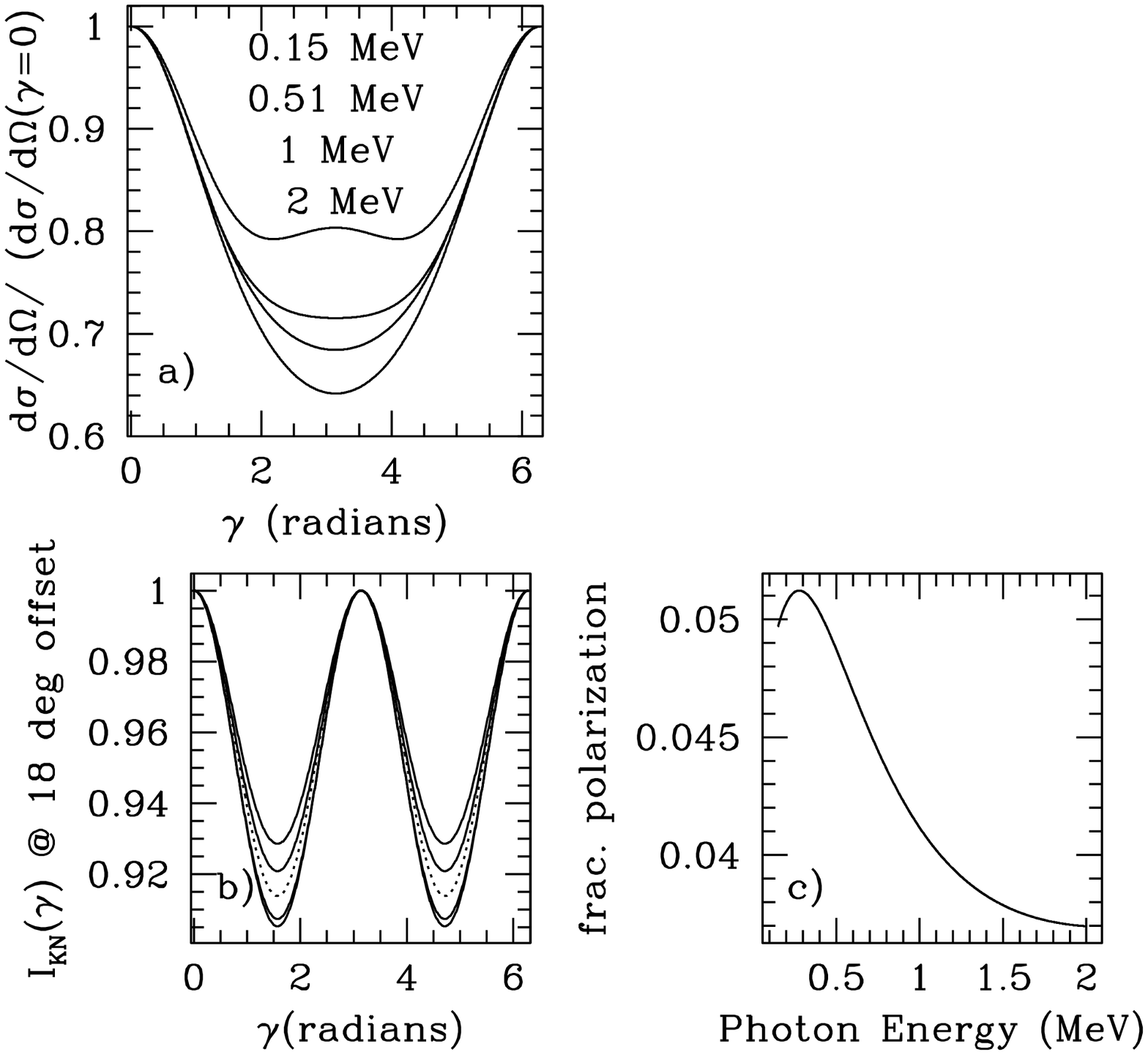 hoffset=-80 voffset=-80}{14.7cm}{21.5cm}
\FigNum{\ref{fig:kn}}
\end{figure}

\clearpage
\pagestyle{empty}
\begin{figure}[htb]
\PSbox{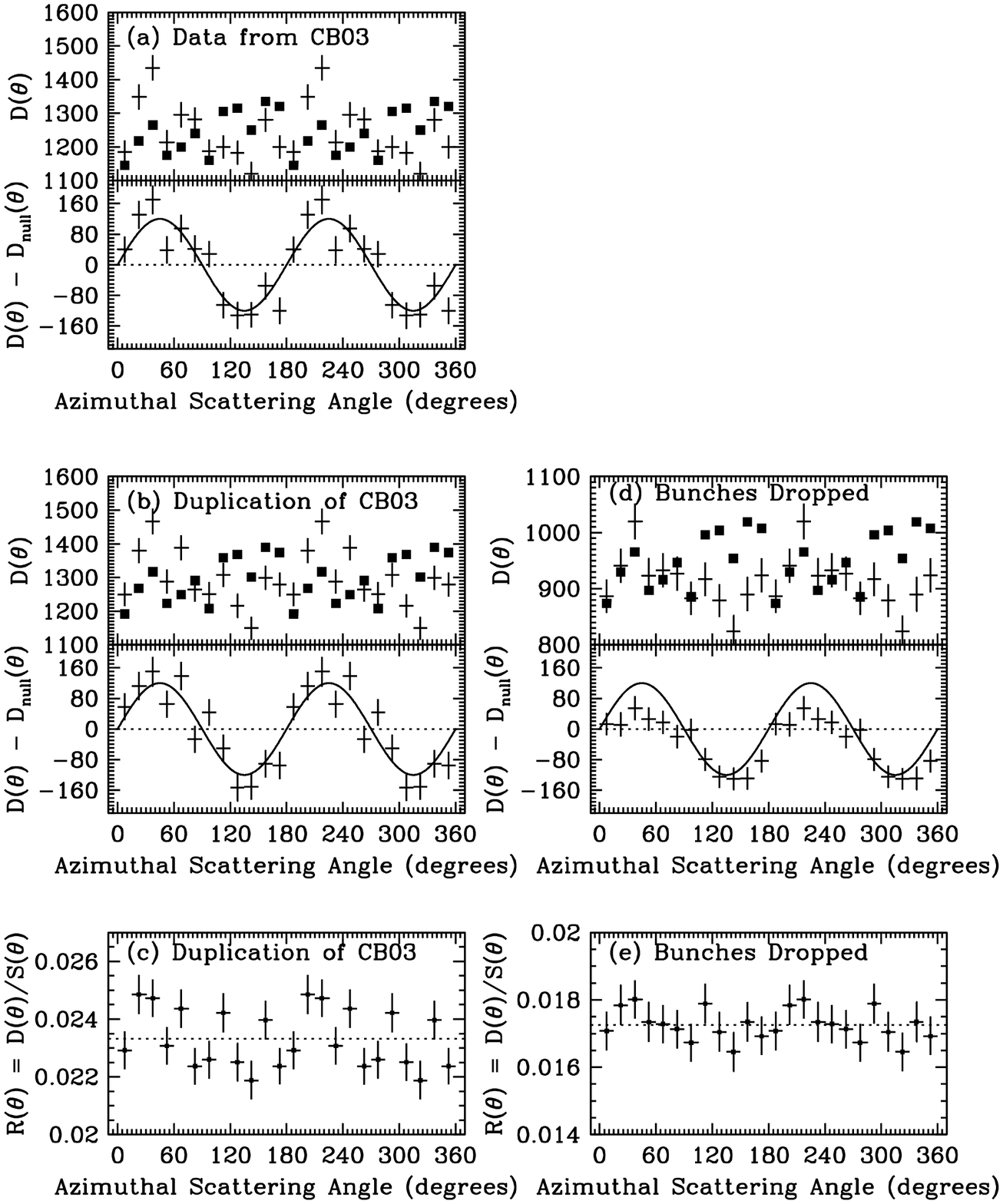  hoffset=-80 voffset=-80}{14.7cm}{21.5cm}
\FigNum{\ref{fig:cb03}}
\end{figure}

\begin{deluxetable}{rrrrrr}
\tablecaption{\label{tab:spex} Detector Physical Parameters}
\tablehead{
\colhead{Detector \#}  & 
\colhead{X} & 
\colhead{Y} &
\colhead{$\Omega_i$} & 
\colhead{$I_i$} \\
\colhead{}  & 
\colhead{(cm)} & 
\colhead{(cm)} &
\colhead{(relative solid angle)} & 
\colhead{(counts)}
}

\startdata
1&   -7.4  &  -3.0  &  3.73 &  9870 \\
2&    7.4  &  -3.0  &   n/a &  n/a \\
3&   15.4  &   5.0  &  1.90 &  8191 \\
4&  -15.4  &   5.0  &  2.68 &  8426 \\
5&    8.2  &  14.0  &  2.52 &  8114 \\
6&   -8.2  &  14.0  &  2.94 &  7379 \\
7&    0.0  &   5.8  &  4.23 &  8377 \\
8&   -5.6  & -14.9  &  2.24 &  8903 \\
9&    5.6  & -14.9  &  1.80 &  7216 \\
\enddata
\end{deluxetable}

\begin{deluxetable}{lcc}
\tablecaption{\label{tab:bg} Double-Count Events and Irreducible Background}
\tablehead{
\colhead{Events}  & 
\colhead{Present work} & 
\colhead{CB03}
}
\startdata
Total Double-Count Events Found  &  8230         & 14916 \\
Coincidences        &  6640\ppm80   & 4488\ppm72 \\
Other Background$^a$ & 760\ppm110  & 588\ppm25  \\  \hline
Double-Count Scattering Events$^b$ & 830\ppm150 & 9840\ppm96 \\
\enddata
\tablecomments{
$^a$In the present work this is the background due to the ``unknown''
mechanism.  In CB03, this is the value labeled ``background scatters''.
$^b$Using a second, independent method which does not require detailed
examination of the different background contributions, we find
910\ppm250 two-detector scattering events, which lends us confidence
in this value. See \S\ref{sec:simple}.
}
\end{deluxetable} 


\begin{thebibliography}{}

\bibitem[\protect\astroncite{{Atteia} {\rm et~al.\/}}{1999}]{atteia99}
{Atteia}, J.-L. and {Bo{\" e}r}, M. and {Hurley}, K., 1999,
\newblock {\em \aaps} { 138}, 421

\bibitem[\protect\astroncite{{Bloom} {\rm et~al.\/}}{2002a}]{bloom02b}
{Bloom}, J.~S., {Kulkarni}, S.~R., \& {Djorgovski}, S.~G., 2002a,
\newblock {\em \aj} { 123}, 1111

\bibitem[\protect\astroncite{{Bloom} {\rm et~al.\/}}{1999}]{bloom99}
{Bloom}, J.~S., {Kulkarni}, S.~R., {Djorgovski}, S.~G., {Eichelberger}, A.~C.,
  {Cote}, P., {Blakeslee}, J.~P., {Odewahn}, S.~C., {Harrison}, F.~A., {Frail},
  D.~A., {Filippenko}, A.~V., {Leonard}, D.~C., {Riess}, A.~G., {Spinrad}, H.,
  {Stern}, D., {Bunker}, A., {Dey}, A., {Grossan}, B., {Perlmutter}, S.,
  {Knop}, R.~A., {Hook}, I.~M., \& {Feroci}, M., 1999,
\newblock {\em \nat} { 401}, 453

\bibitem[\protect\astroncite{{Bloom} {\rm et~al.\/}}{2002b}]{bloom02}
{Bloom}, J.~S., {Kulkarni}, S.~R., {Price}, P.~A., {Reichart}, D., {Galama},
  T.~J., {Schmidt}, B.~P., {Frail}, D.~A., {Berger}, E., {McCarthy}, P.~J.,
  {Chevalier}, R.~A., {Wheeler}, J.~C., {Halpern}, J.~P., {Fox}, D.~W.,
  {Djorgovski}, S.~G., {Harrison}, F.~A., {Sari}, R., {Axelrod}, T.~S.,
  {Kimble}, R.~A., {Holtzman}, J., {Hurley}, K., {Frontera}, F., {Piro}, L., \&
  {Costa}, E., 2002b,
\newblock {\em \apjl} { 572}, L45

\bibitem[\protect\astroncite{{Bloser} {\rm et~al.\/}}{2003}]{bloser03}
{Bloser}, P.~F., {Hunter}, S.~D., {Depaola}, G.~O., \& {Longo}, F., 2003,
\newblock in {\em Proc. SPIE Vol. 5165, "X-ray and Gamma-Ray Instrumentation
  for Astronomy XIII"},
\newblock astro-ph/0308331

\bibitem[\protect\astroncite{{Coburn} \& {Boggs}}{2003}]{coburn03}
{Coburn}, W. \& {Boggs}, S.~E., 2003,
\newblock {\em \nat} { 423}, 415

\bibitem[\protect\astroncite{{Eichler} \& {Levinson}}{2003}]{eichler03}
{Eichler}, D. \& {Levinson}, A., 2003,
\newblock {\em \apjl} { 596}, 147

\bibitem[\protect\astroncite{{Frail}}{2003}]{frail03}
{Frail}, D.~A., 2003,
\newblock {\em GRB Circular Network} 2280

\bibitem[\protect\astroncite{{Granot}}{2003}]{granot03}
{Granot}, J., 2003,
\newblock {\em \apjl},
\newblock in press, astro-ph/0306322

\bibitem[\protect\astroncite{{Hjorth} {\rm et~al.\/}}{2003}]{hjorth03}
{Hjorth}, J., {Sollerman}, J., {M{\o}ller}, P., {Fynbo}, J.~P.~U., {Woosley},
  S.~E., {Kouveliotou}, C., {Tanvir}, N.~R., {Greiner}, J., {Andersen}, M.~I.,
  {Castro-Tirado}, A.~J., {Castro Cer{\' o}n}, J.~M., {Fruchter}, A.~S.,
  {Gorosabel}, J., {Jakobsson}, P., {Kaper}, L., {Klose}, S., {Masetti}, N.,
  {Pedersen}, H., {Pedersen}, K., {Pian}, E., {Palazzi}, E., {Rhoads}, J.~E.,
  {Rol}, E., {van den Heuvel}, E.~P.~J., {Vreeswijk}, P.~M., {Watson}, D., \&
  {Wijers}, R.~A.~M.~J., 2003,
\newblock {\em \nat} { 423}, 847

\bibitem[\protect\astroncite{{Hurley} {\rm et~al.\/}}{2003}]{hurley03}
{Hurley}, K., {Cline}, T., {Smith}, D.~M., {Lin}, R.~P., {McTiernan}, J.,
  {Schwartz}, R., {Wigger}, C., {Hajdas}, W., {Zehnder}, A., {Mitrofanov}, I.,
  {Charyshnikov}, S., {Grinkov}, V., {Kozyrev}, A., {Litvak}, M., {Sanin}, A.,
  {Boynton}, W., {Fellows}, C., {Harshman}, K., {Shinohara}, C., {Starr}, R.,
  {Mazets}, E., {Golenetskii}, S., {von Kienlin}, A., {Lichti}, G., \& {Rau},
  A., 2003,
\newblock {\em GRB Circular Network} { 2281}, 1

\bibitem[\protect\astroncite{{Lazzati} {\rm et~al.\/}}{2003}]{lazzati03}
{Lazzati}, D., {Rossi}, E., {Ghisellini}, G., \& {Rees}, M.~J., 2003,
\newblock {\em \aa}

\bibitem[\protect\astroncite{{Lin} {\rm et~al.\/}}{2002}]{rhessi}
{Lin}, R.~P., {Dennis}, B.~R., {Hurford}, G.~J., {Smith}, D.~M., {Zehnder}, A.,
  {Harvey}, P.~R., {Curtis}, D.~W., {Pankow}, D., {Turin}, P., {Bester}, M.,
  {Csillaghy}, A., {Lewis}, M., {Madden}, N., {van Beek}, H.~F., {Appleby}, M.,
  {Raudorf}, T., {McTiernan}, J., {Ramaty}, R., {Schmahl}, E., {Schwartz}, R.,
  {Krucker}, S., {Abiad}, R., {Quinn}, T., {Berg}, P., {Hashii}, M.,
  {Sterling}, R., {Jackson}, R., {Pratt}, R., {Campbell}, R.~D., {Malone}, D.,
  {Landis}, D., {Barrington-Leigh}, C.~P., {Slassi-Sennou}, S., {Cork}, C.,
  {Clark}, D., {Amato}, D., {Orwig}, L., {Boyle}, R., {Banks}, I.~S., {Shirey},
  K., {Tolbert}, A.~K., {Zarro}, D., {Snow}, F., {Thomsen}, K., {Henneck}, R.,
  {Mchedlishvili}, A., {Ming}, P., {Fivian}, M., {Jordan}, J., {Wanner}, R.,
  {Crubb}, J., {Preble}, J., {Matranga}, M., {Benz}, A., {Hudson}, H.,
  {Canfield}, R.~C., {Holman}, G.~D., {Crannell}, C., {Kosugi}, T., {Emslie},
  A.~G., {Vilmer}, N., {Brown}, J.~C., {Johns-Krull}, C., {Aschwanden}, M.,
  {Metcalf}, T., \& {Conway}, A., 2002,
\newblock {\em \solphys} { 210}, 3

\bibitem[\protect\astroncite{{Lyutikov} {\rm et~al.\/}}{2003}]{lyutikov03}
{Lyutikov}, M., {Pariev}, V.~I., \& {Blandford}, R.~D., 2003,
\newblock {\em \apj},
\newblock submitted, astro-ph/0305410

\bibitem[\protect\astroncite{{Matsumiya} \& {Ioka}}{2003}]{matsumiya03}
{Matsumiya}, M. \& {Ioka}, K., 2003,
\newblock {\em \apjl} { 595}, L25

\bibitem[\protect\astroncite{{McConnell} {\rm et~al.\/}}{2002}]{mcconnell02}
{McConnell}, M.~L., {Ryan}, J.~M., {Smith}, D.~M., {Lin}, R.~P., \& {Emslie},
  A.~G., 2002,
\newblock {\em \solphys} { 210}, 125

\bibitem[\protect\astroncite{{Nakar} {\rm et~al.\/}}{2003}]{nakar03}
{Nakar}, E., {Piran}, T., \& {Waxman}, E., 2003,
\newblock astro-ph/03407290

\bibitem[\protect\astroncite{{Smith}}{1998}]{smithself02}
{Smith}, D.~M., 1998,
\newblock {\em \mnras} 301

\bibitem[\protect\astroncite{{Smith} {\rm et~al.\/}}{2002}]{spex}
{Smith}, D.~M., {Lin}, R.~P., {Turin}, P., {Curtis}, D.~W., {Primbsch}, J.~H.,
  {Campbell}, R.~D., {Abiad}, R., {Schroeder}, P., {Cork}, C.~P., {Hull},
  E.~L., {Landis}, D.~A., {Madden}, N.~W., {Malone}, D., {Pehl}, R.~H.,
  {Raudorf}, T., {Sangsingkeow}, P., {Boyle}, R., {Banks}, I.~S., {Shirey}, K.,
  \& {Schwartz}, R., 2002,
\newblock {\em \solphys} { 210}, 33

\bibitem[\protect\astroncite{{Stanek} {\rm et~al.\/}}{2003}]{stanek03}
{Stanek}, K.~Z., {Matheson}, T., {Garnavich}, P.~M., {Martini}, P., {Berlind},
  P., {Caldwell}, N., {Challis}, P., {Brown}, W.~R., {Schild}, R.,
  {Krisciunas}, K., {Calkins}, M.~L., {Lee}, J.~C., {Hathi}, N., {Jansen},
  R.~A., {Windhorst}, R., {Echevarria}, L., {Eisenstein}, D.~J., {Pindor}, B.,
  {Olszewski}, E.~W., {Harding}, P., {Holland}, S.~T., \& {Bersier}, D., 2003,
\newblock {\em \apjl} { 591}, L17

\end{thebibliography}
\end{document}